\documentclass[print]{elsarticle}
\usepackage{lineno,hyperref}
\usepackage{graphicx}
\usepackage[linesnumbered,ruled,lined]{algorithm2e}
\usepackage{float}
\usepackage{tikz}
\usepackage{epstopdf}
\usepackage{amssymb}
\usepackage{amsmath}
\usepackage{subcaption}
\usepackage{footnote}
\usepackage{array}
\usepackage{cases}
\usepackage{multirow}
\usepackage{siunitx}
\usepackage{booktabs}
\usepackage{rotating}
\usepackage{threeparttable}
\journal{International Journal of Heat and Mass Transfer}
\usepackage{makecell}

\definecolor{chiColor}{rgb}{255,0,0}

\begin{document}
%
%
\begin{frontmatter}
		\title{A GPU-Accelerated Fully Coupled Fluid–Solid–Thermal SPH Solver for 
    Industrial Gearboxes: Application to Lubricant Flow and Heat Transfer in a Bevel–Helical Reducer}
		\author[myfirstaddress,mysecondaryaddress]{Yongchuan Yu }
        \ead{yongchuan.yu@tum.de /yongchuan.yu@einsimo.com}
        
        \author[mythirdaddress,mysecondaryaddress]{Dong Wu\corref{mycorrespondingauthor} }
		\ead{dong.wu@tum.de /dong.wu@einsimo.com}

		\author[myfirstaddress]{Oskar J. Haidn}
		\ead{oskar.haidn@tum.de}
        
        \author[mythirdaddress]{Xiangyu Hu}
        \ead{xiangyu.hu@tum.de}

		\address[myfirstaddress]{Chair of Space Propulsion, 
			Technical University of Munich, 85521 Ottobrunn, Germany}
		\address[mysecondaryaddress]{EinSIMO GmbH, 80687 München, Germany}
            \address[mythirdaddress]{Chair of Aerodynamics and Fluid Mechanics, 
			Technical University of Munich, 85748 Garching bei München, Germany}
		\cortext[mycorrespondingauthor]{Corresponding author. }
\begin{abstract}

This study presents a GPU-accelerated, fully coupled fluid–solid–thermal 
Smoothed Particle Hydrodynamics (SPH) framework for high-fidelity analysis of 
splash-lubricated gearboxes. A series of thermo–fluid simulations of a 
bevel–helical gear reducer were conducted by varying shaft speed, oil immersion 
depth, and lubricant viscosity to evaluate their influence on splash dynamics, 
churning losses, and lubricant temperature rise. The results show that 
churning losses increase by nearly an order of magnitude as the speed rises 
from 150 to 600~$\mathrm{rad/s}$, while the corresponding lubricant temperature 
rise becomes approximately three to four times smaller. Variations in immersion 
depth and viscosity adjust the heating rate only modestly—typically within 
10–20\%—with their influence reversing between low- and high-speed regimes. 
The GPU backend provides a 7–9× speedup over a high-performance desktop CPU, 
enabling multi-million-particle, full-gearbox thermo–fluid simulations without 
specialized hardware. These findings demonstrate the feasibility of high-fidelity 
thermal analysis of industrial gearboxes and provide quantitative insight into 
the coupled splash, churning, and heat-transfer mechanisms that govern gearbox 
thermal performance.

\end{abstract}

\end{frontmatter}
%
%
\section{Introduction}\label{sec:introduction}
Gear transmissions are fundamental components in a wide range of engineering systems, 
including automotive, rail, wind turbines, and aviation sectors 
\cite{Industrial-automotive--abel2006engine, 
Industrial-railway--deng2020lubrication, Industrial-wind-trubine--ragheb2010wind, 
Industrial-marine--song2015dynamic, Industrial-aviation--hu2021influence}. 
As energy efficiency and environmental regulations become increasingly stringent, 
there is a growing demand for gearbox designs that minimize power losses and operating 
temperatures while enhancing reliability and service life \cite{overall--concli2023latest}. 
Specifically, load-independent power losses, consisting primarily of churning, 
windage, and squeezing effects, represent a substantial portion of the overall energy 
dissipation in gear transmission systems, significantly influencing their operational 
efficiency and thermal behavior \cite{overall-experiment--mastrone2020oil, 
overall--liu2018detailed}.
To better understand and mitigate such losses, and thereby achieve more accurate prediction 
and control of the thermal behavior inside the gearbox, numerous experimental investigations 
have been carried out. For instance,
Hartono \cite{experiment-piv--hartono2013piv} employed 2D particle image velocimetry 
(PIV) in a transparent dip-lubricated gearbox to examine lubricant flow under varying 
oil depths and gear speeds.
Mastrone\cite{overall-experiment--mastrone2020oil} combined PIV measurements with torque 
analysis to quantify oil distribution and validate numerical simulations.
Through a dedicated test, 
Laruelle \cite{experiment-spiral-bevel-gear--laruelle2017experimental}demonstrated 
significant insights into churning losses of spiral bevel gears, 
leading to improved empirical predictions beyond existing models.
However, direct experimental characterization of internal temperature fields, 
lubricant thermal stratification, or heat-dissipation paths remains extremely challenging.
A notable attempt to address gearbox thermal behavior at a system level is the 
reduced-order thermo-fluid model by Yazdani\cite{Reduced-order-model-yazdani2015prediction}.
By combining multiscale analysis, regime mapping, and simplified heat-flow formulations, 
their work provides physical insight into dominant thermo-fluid mechanisms.
Yet these quasi-steady models rely on simplifying assumptions and cannot capture the transient, 
splash-driven thermal behavior of real gearboxes.

Based on these experimental foundations, computational fluid dynamics (CFD) has been 
extensively used to model splash phenomena, churning mechanisms, 
and thermal transport in gearboxes. 
Early steady‐state Reynolds-Averaged Navier–Stokes
(RANS) simulations assessed oil‐guide effects and lubricant transport 
in spiral bevel gears \cite{CFD-unsteady--jiang2019influences}.
Subsequent VOF-based CFD studies have broadly examined splash formation \cite{CFD-VOF-MRF--lu2020cfd}, 
free-surface films \cite{CFD-MRF-Plate--hildebrand2022cfd}, 
jet-injection splash patterns \cite{CFD-transient-VOF--keller2020cfd}, 
and shroud-geometry optimization \cite{CFD-heat--qian2024optimization} for 
reducing churning torque in various gearbox configurations.
To incorporate thermal effects, Hu et al. \cite{Thermal-VOF-hu2024study} coupled 
VOF simulations with a steady thermal FE model to predict gear-surface temperatures 
under oil-jet lubrication, while Qiao et al. \cite{CFD-Heat-flow-coupled-qiao2024novel} 
developed a dynamic heat-flow coupled spray-lubrication model validated by infrared thermal imaging.
To improve transient fidelity, 
dynamic‐mesh, sliding‐mesh, and immersed‐solid strategies were compared, demonstrating that dynamic 
mesh best preserves element quality and temperature evolution for churning‐loss 
analysis \cite{CFD-moving-mesh--li2022comparative}.  
However, mesh‐based approaches remain constrained by mesh distortion and high computational cost, 
particularly in highly dynamic, free‐surface flows in complex gear geometries, 
making full gearbox-scale thermal prediction difficult.

Meshless methods have emerged as viable alternatives due to their ability to handle large 
deformations and moving interfaces without the burden of mesh generation \cite{SPH--liu2019numerical}. 
Among them, lattice‐Boltzmann–LES frameworks have provided fine‐scale insights into 
vortex dynamics, heat distribution, and viscosity‐dependent mixing in two‐stage 
transmissions \cite{CFD-LBM-heat--zheng2025investigate}. 
Moving‐Particle Semi‐Implicit (MPS) simulations have quantified splash lubrication 
and churning losses in two‐stage rail reducers \cite{MPS-rail-vehicle--liu2023research} 
and spiral bevel gearboxes \cite{MPS-spiral-bevel--shen2023research} without altering gear geometry. 
Furthermore, MPS has been used to predict flow and temperature fields in accessory gearboxes 
through coupling with a thermal-network model \cite{MPS-Thermal-lin2025numerical}.
In parallel, Smoothed Particle Hydrodynamics (SPH) was adopted to map meshing‐impact 
stresses \cite{SPH-impact--imin2014stress} at an early stage, then extended to visualize 
oil‐bath flows in dip‐lubricated gearboxes, showing good agreement with PIV 
measurements \cite{SPH-ji2018numerical}. 
Liu et al. later extended SPH to simulate oil distribution and churning gear power losses 
under dip‐lubrication \cite{SPH--liu2019numerical}. Hybrid SPH–FEM couplings have captured 
transient rack‐and‐pinion impact dynamics \cite{SPH-FEM--di2023meshing}, while SPH integrated with 
Taguchi optimization has identified key parameters for improved lubricant distribution in rail‐vehicle 
gearboxes \cite{SPH-optimization--yang2025flow}.
Recent SPH developments have introduced fully coupled fluid–solid–thermal models, 
enabling more realistic simulations of lubrication and cooling 
processes \cite{SPH-fluid-solid-heat--shi2025sph}.
Building on these advances, SPH has also been applied to lubricated gear meshing with explicit
fluid–solid–heat coupling, resolving local temperature variation under oil-bath and
oil-jet cooling conditions \cite{SPH-Thermal-shi2025sph}.

Despite these advances, ultra-high shaft speeds, where tip velocities approach the same 
order of magnitude as the speed of sound, impose prohibitively tiny time step sizes, 
making even meshless simulations computationally expensive under industrial-scale conditions.
When such high-speed conditions are combined with fine spatial resolutions or large 
particle counts, the computational burden escalates sharply.
Thus far, few studies have comprehensively addressed the coupled phenomena of transient splash distribution, 
churning losses, and heat transfer in complex gearboxes operating under such extreme conditions, 
while also maintaining tractable runtimes. 
The all-in-one multiphysics simulations required for this remain computationally demanding, 
with even a single case often taking days, limiting their practicality for routine design optimization.
To address these challenges, we enhance our open-source SPH library -  
SPHinXsys\cite{method--ZHANG/SPHinXsys, method--zhang2021SPHinXsys} by integrating a 
SYCL-based (a Khronos-standard C++ heterogeneous programmingmodel) GPU 
acceleration framework, achieving even a tenfold runtime improvement over the baseline.
This work develops a fully coupled SPH solver that simultaneously 
captures free-surface lubrication flow, fluid–solid interaction, and thermal transport. 
A contact-factor metric is further introduced to quantify the instantaneous lubricant 
coverage on gear surfaces and its implications for flow distribution and heat transfer.
The solver is validated against benchmark experiments on a type C-PT FZG gear set, 
followed by a multi-parameter study evaluating the effects of immersion depth, 
rotational speed, and oil viscosity in a complex industrial level Bevel–Helical gear reducer.
The results provide new insights into the interplay between splash dynamics, churning loss, 
and heat transfer, while offering an efficient computational toolkit for high-performance 
gearbox design and optimization.

The remainder of this paper is organized as follows.
Section \ref{sec:methodology} outlines the governing equations and numerical formulations 
implemented in the GPU-accelerated SPHinXsys framework.
Section \ref{sec:validation} first compares simulated churning losses with experimental 
data for a classical type C-PT FZG gearbox, and then verifies the heat transfer modeling 
against the analytical solution of transient conduction between parallel plates.
Section \ref{sec:Bevel-Helical-gearbox-simulation} presents a multi-parameter study of a 
bevel–helical industrial gearbox, analyzing the influence of operating conditions on both 
lubrication flow and thermal behavior, and evaluates the computational performance of the 
SYCL-based GPU implementation against traditional CPU execution.
Section \ref{sec:conclusion} concludes the work.
The code of the present method is available in the open-source SPHinXsys 
repository \cite{method--ZHANG/SPHinXsys, method--zhang2021SPHinXsys} 
at \url{https://www.sphinxsys.org}.

%
%
\section{Methodology}
\label{sec:methodology}
In this section, 
we summarize the methodologies employed in this paper, 
including the governing equations and special numerical algorithms to 
facilitate the calculation of complex gearbox simulation. 
\subsection{Governing Equations}
\label{subsec:governing-equations} 
In this paper, viscous, incompressible flow acting on rigid bodies is considered.  
The mass and momentum conservation equations can be written as:
\begin{equation}
\begin{cases}
	\frac{\text{d} \rho}{\text{d} t}  =  - \rho \nabla \cdot \mathbf{v} \\
	 \frac{\text{d} \mathbf{v}}{\text{d} t}  = \frac{1}{\rho} (- \nabla p+\eta\nabla^2\mathbf{v} )  + \mathbf{g}+\mathbf{f}^{s:p}+\mathbf{f}^{s:\nu}
 \end{cases},
 \label{eq:mass and momentum}
\end{equation}
where $\frac{\text{d}}{\text{d} t} = \frac{\partial}{\partial t} + \mathbf{v} \cdot \nabla$ 
stands for the material derivative,  
while $\rho$ the density, $\mathbf{v}$ the velocity, $p$ the pressure, 
$\mathbf{g}$ the gravity, $\eta$ the dynamic viscosity, 
$\mathbf{f}^{s:p}$ the pressure force acting on fluid from the rigid wall 
and $\mathbf{f}^{s:\nu}$ the viscous force acting on the fluid from the rigid wall.  
No explicit turbulence model is used in this study. 
Instead, the weakly compressible SPH formulation acts as an implicit large-eddy simulation (iLES), 
where numerical dissipation associated with kernel smoothing provides sub-grid 
regularization \cite{SPH-iLES-ferrand2013unified, SPH-iLES-violeau2016smoothed, SPH-iLES-di2017smoothed}

To close the system of equations \ref{eq:mass and momentum}, 
an artificial isothermal equation of state (EoS) 
\begin{equation}
p = {c^0}^2(\rho - \rho^0), 
\label{eq2}
\end{equation}
is adopted with the weakly-compressible assumption 
\cite{method--monaghan1994simulating, method--morris1997modeling}. Here, $\rho^0$ denotes the 
initial reference density, while the $c^0 = 10 \times U_{max}$ is the artificial 
speed of sound relating to the maximum anticipated flow speed $U_{max}$.  

Since one of our main goals is to study the temperature distribution inside the gearbox, 
the heat transfer equation must also be taken into account.
As convective heat transfer is already embedded in the Lagrangian form of the governing 
equations by tracking particle temperatures, the calculation of temperature variation 
can be limited to the thermal diffusion component alone \cite{SPH-Heat-Transfer-Industral-ng2019fluid,
SPH-Heat-Transfer-Industral-abdolahzadeh2021thermal, SPH-Heat-Transfer-Industral-hosain2019smoothed}.
The equation of thermal diffusion can be written as:
\begin{equation}
    \rho\frac{\text{d}U}{\text{d}t}=\nabla \cdot (k\nabla T),
    \label{eq:thermal-diffusion-U}
\end{equation}
where $U$ is the internal energy, $k$ the thermal conductivity coefficient. 
An equivalent formulation can also be written in terms of temperature: 
\begin{equation}
    \frac{\text{d}T}{\text{d}t}=\nabla \cdot (\alpha \nabla T),
    \label{eq:thermal-diffusion-T}
\end{equation}
where $T$ is the temperature of the particle, $\alpha=\frac{k}{\rho c_p}$ 
is the thermal diffusivity. Here the $c_p$ and $k$ denote 
the specific heat capacity and thermal conductivity, respectively.
\subsection{Physics-driven particle relaxation} 
\label{subsec:particle-relaxation}
A key requirement for particle-based simulations is the generation of an isotropic, 
body-fitted particle distribution, particularly crucial for handling complex geometries 
such as the multi-stage gearbox discussed in this study. To fulfill this requirement, 
a physics-driven relaxation process, 
as detailed in Refs. \cite{method--yujiezhu2021, method--yu_level-set,
method--zhao2025physics}, is employed. 
In this pre-processing approach, the geometry boundary is implicitly defined 
by the zero level-set contour of a signed distance field $\phi(x,y,z)$.
\begin{equation}
    \Gamma = \left\{(x,y,z)|\phi(x,y,z,t) = 0\right\}.
    \label{eq:zero-level-set}
\end{equation}
Negative values of the level-set function correspond to the interior of the geometry, 
while positive values represent the exterior region.

Once the geometry surface has been defined, an initial set of discrete particles 
is generated on a predefined lattice embedded within the target geometry. 
These particles are subsequently driven toward an isotropic and boundary-conforming 
configuration through a relaxation process governed by a constant background pressure field.
\begin{equation}
    \frac{\text{d} \mathbf{v}}{\text{d}t}=\mathbf{F}_p,
    \label{eq:physics-relax}
\end{equation}
where $\mathbf{v}$ is the advection velocity, $\mathbf{F}_p$ represents the 
accelerations induced by the background pressure field.
The right-hand side of Eq. \ref{eq:physics-relax} can be reformulated using 
the SPH discretization.
\begin{equation}
    \boldsymbol{F}_{p, i}=-\frac{2}{m_{i}} \sum_{j} V_{i} V_{j} p_{0} \nabla_{i} W_{i j},
    \label {eq:physics-relax-SPH}
\end{equation}
where $m$ is the particle mass, $V$ the particle volume, 
$p_0$ the constant background pressure and $\nabla_{i} W_{i j}$ 
represents the gradient of the kernel function 
$W(\left |\mathbf{r}_{ij} \right |, h)$ with respect to particle $i$. 
Here, $\mathbf{r}_{ij}=\mathbf{r}_i-\mathbf{r}_j$ and $h$ stands 
for the directed particle spacing and smoothing length. 
To compensate for the kernel deficiency near geometric boundaries, 
a static confinement method based on the background level-set field is employed, 
as described in Ref \cite{method--yu_level-set}. 
Meanwhile, for more information on how to obtain better zero-order consistency for
physics-driven relaxation methods of multi-body systems, 
please refer to Ref \cite{method--zhao2025physics}.

The allowable time-step $\Delta t$ is determined by the stability criterion for body forces:
\begin{equation}
   \Delta t\le 0.25\sqrt{\frac{h}{\left | \mathrm{d} \mathbf{v} /\mathrm{d} t  \right | } }. 
\end{equation}
The position-updating method of particles is defined as
\begin{equation}
    \mathbf{r} ^{n+1}=\mathbf{r}^{n} + \mathrm{d}\mathbf{r} = \mathbf{r}^{n}+\frac{1}{2}  \mathbf{F}_p^n
\Delta t^2. 
\end{equation}.
\subsection{Riemann based WSPH solvers} \label{subsec:riemann-sph} 
In SPHinXsys, the weakly compressible SPH solution is adopted.
According to Ref.\cite{method--zhang2017weakly}, 
the mass and momentum conservation equations in Eq.\ref{eq:mass and momentum} 
are reformulated using a Riemann-based SPH discretization:
\begin{equation}
    \begin{cases}
		\frac{\text{d} \rho_i}{\text{d} t} = 2\rho_i \sum_jV_j(\mathbf{v}_{i}-\mathbf{v}^{\ast})  \nabla_i W_{ij} \\
		\frac{\text{d} \mathbf{v}_i}{\text{d} t}  = - \frac{2}{m_i}\sum_j V_iV_j p^{\ast}  \nabla_i W_{ij}+
  \frac{2}{m_i}\sum_j V_iV_j \frac{\eta_{ij}\mathbf{v_{ij}}}{r_{ij}}\frac{\partial W_{ij}}{\partial r_{ij}}  
  +\mathbf{g}+\mathbf{f}_i^{s:p}+\mathbf{f}_i^{s:\nu}
	\end{cases},
 \label{eq:riemann-mass-momentum}
\end{equation}
where $V$ is particle volume, $m$ is the particle mass, 
$\mathbf{v}_{ij} = \mathbf{v}_{i} - \mathbf{v}_{j}$ is the relative velocity, 
$\mathbf{r}$ is the particle position, 
$r_{ij}=\left | \mathbf{r}_i-\mathbf{r}_j\right |$ is the particle distance,
 $\nabla_i W_{ij}=\mathbf{e}_{ij} \frac{\partial W_{ij}}{\partial r_{ij}}$ is the gradient of the
 kernel function of particle $i$, and $\mathbf{v}^\ast$ and  $p^\ast$ are the 
 solutions of inter-particle Riemann problem along the unit 
 vector $\mathbf{e}_{ij} = \frac{\mathbf{r}_{ij}}{r_{ij}}$\cite{method--zhang2017weakly}. 

The reconstruction of the initial left and right states in the inter-particle Riemann problem is given by:
\begin{equation}
    \begin{cases}
	(\rho_L, U_L, P_L) = (\rho_i,- \mathbf{v}_i\cdot \mathbf{e}_{ij}, p_i)\\
	(\rho_R, U_R, P_R) = (\rho_j,-\mathbf{v}_j\cdot \mathbf{e}_{ij}, p_j)
	\end{cases}.
 \label{eq:initial-left-right-states}
\end{equation}
where the subscripts $L$ and $R$ are the left and right states, 
respectively \cite{method--zhang2017weakly}. 
The Riemann solution incorporating a dissipation limiter, 
as proposed in Ref.\cite{method--zhang2017weakly}, can then be obtained as:
\begin{equation}
    \begin{cases}
	 U^\ast = \bar U + \frac{p_L-p_R}{c(\rho_L+\rho_R)} \\
	 p^\ast = \bar P + \frac{\rho_L \rho_R \beta (U_L-U_R)}{\rho_L+\rho_R}
	\end{cases},
 \label{eq:riemann-solution}
\end{equation}
Here, the dissipation limiter is defined as $\beta = min[3max(U_L-U_R, 0), c^0]$.
This formulation ensures that no artificial dissipation is introduced in regions 
undergoing expansion, while adaptively regulating the level of numerical dissipation 
in compressive regions, as described in Ref.\cite{method--zhang2017weakly}.
The value of $\mathbf{v}^\ast$ is subsequently obtained from:
\begin{equation}
    \mathbf{v}^{*}=\left(U^{*}-\frac{\rho_{L} U_{L}+\rho_{R} U_{R}}{\rho_{L}+\rho_{R}}\right) \mathbf{e}_{i j}+\frac{\rho_{i} \mathbf{v}_{i}+\rho_{j} \mathbf{v}_{j}}{\rho_{i}+\rho_{j}}.
\label{eq:v-star}
\end{equation}

\subsection{Density re-initialization} \label{subsec:density-reinitialization} 
Based on the Riemann-based SPH formulation described in Section\ref{subsec:riemann-sph}, 
a density reinitialization scheme \cite{method--zhang2020dual} is employed to enhance the stability of the density field updated through the continuity equation \ref{eq:riemann-mass-momentum}. 
At the start of each new time step, the fluid density in regions identified as 
free-surface flow is reinitialized according to:

\begin{equation}
    \rho_i= max(\rho^\ast, \rho^0 \frac{\sum W_{ij}}{\sum W^0_{ij}}).
    \label{eq:density-reinitialized}
\end{equation}

where $\rho^\ast$ stands for the density before reinitialization, 
and the superscript $0$ refers to the initial states.

In the absence of a free surface, Eq.\ref{eq:density-reinitialized} is modified accordingly as:

\begin{equation}
    \rho_i=\rho^0 \frac{\sum W_{ij}}{\sum W^0_{ij}}. 
    \label{eq:density-light-phase}
\end{equation}

\subsection{Boundary treatment} \label{subsec:boundary} 
To impose solid wall boundary conditions, dummy particles are introduced following 
the approach proposed in Refs.\cite{method--adami2012generalized, method--zhang2017weakly}. 
The interaction between fluid and wall particles is modeled by solving a one-sided Riemann 
problem oriented along the wall-normal direction \cite{method--zhang2017weakly}. 
The left state of this one-sided Riemann problem is defined as:
\begin{equation}
    (\rho_L, U_L, P_L) = (\rho_f, -\mathbf{n}_w \cdot \mathbf{v}_f, P_f),
    \label{eq:wall-left-state}
\end{equation}
where the subscript $f$ is for fluid particles, the $\mathbf{n}_w$ is the local wall 
normal direction. 
While the right states as follows:
\begin{equation}
    \begin{cases}
	 U_R = - U_L + 2 u_w \\
	 P_R = P_L + \rho_f \mathbf{g} \cdot \mathbf{r}_{f_w} 
	\end{cases},
 \label{eq:wall-right-state}
\end{equation}
where the $u_w$ is the velocity of the wall, 
the $\mathbf{r}_{f_w} = \mathbf{r}_w - \mathbf{r}_f$ and 
the density of the right state could be obtained by solving the artificial EoS.
Therefore, the pressure force and the viscous force exerted by the wall on the fluid 
can be written as:
\begin{equation}
    \begin{cases}
		\mathbf{f}_i^{s:p} = - \frac{2}{m_i}\sum_w V_i V_w p^\ast \nabla_i W_{iw} \\
		\mathbf{f}_i^{s:\nu}  = \frac{2}{m_i}\sum_w V_iV_w \frac{2 \eta_{i} \eta_{w}} 
            {\eta_{i} + \eta_{w}} \frac{\mathbf{v}_i - \mathbf{v}_w}{r_{iw}} 
            \frac{\partial W_{iw}}{\partial r_{iw}} 
	\end{cases},
 \label{eq:wall-pressure-viscous}
\end{equation}
Here the subscript $i$ is for fluid particles, and $w$ is for wall particles. The intermediate pressure value $p^\ast$ is obtained as:
\begin{equation}
    p^{*}=\frac{\rho_fp_w+\rho_wp_f}{\rho_f+\rho_w}.
    \label{eq:wall-p-star}
\end{equation}

\subsection{Thermal diffusion} \label{subsec:thermal-diffusion} 
According to the method outlined in Ref. \cite{method--cleary1999conduction}, 
thermal diffusion equations 
Eq. \ref{eq:thermal-diffusion-U} and Eq. \ref{eq:thermal-diffusion-T} can be discretized as:
\begin{equation}
    \frac{\text{d}U_i}{\text{d}t} = \frac{1}{m_i}(\sum_j V_iV_j\frac{4k_i k_j}{k_i + k_j} \frac{T_{ij}}{r_{ij}}\frac{\partial W_{ij}}{\partial r_{ij}})
    \label{eq:Thermal-U-SPH}
   \end{equation}
and
   \begin{equation}
    \frac{\text{d}T_i}{\text{d}t} =\alpha_i (\sum_j V_j\frac{4k_j}{k_i + k_j} \frac{T_{ij}}{r_{ij}}\frac{\partial W_{ij}}{\partial r_{ij}}),
     \label{eq:Thermal-T-SPH}
\end{equation}
Here, the subscript $j$ denotes the neighboring fluid particles of particle $i$, 
$U$ represents the internal energy, and $k$ is the thermal conductivity.
The thermal diffusivity is given by $\alpha = k/(\rho C_p)$, 
where the $\rho$ is the density and $C_p$ is the specific heat capacity,  
while $T_{ij}$ is equals to $T_i - T_j$. 
For the heat exchange between the fluid and the gears or gearbox, 
Equation \ref{eq:Thermal-T-SPH} can be rewritten as:
\begin{equation}
    \frac{\text{d}T_i}{\text{d}t} =\frac{1}{\rho C_p} (\sum_j V_j\frac{4k_i k_j}{k_i + k_j} \frac{T_{ij}}{r_{ij}}\frac{\partial W_{ij}}{\partial r_{ij}} 
    + \sum_n V_n\frac{4k_i k_n}{k_i + k_n} \frac{T_{in}}{r_{in}}\frac{\partial W_{in}}{\partial r_{in}}),
     \label{eq:Thermal-T-SPH-with-wall}
\end{equation}
here the $n$ denotes the neighboring wall particles in the cut-off range of particle $i$,
other parameters with subscript $n$ can be compared to neighboring particles in the fluid.

The present thermal model does not include the viscous dissipation term 
\begin{equation}
    \Phi=2 \mu S_{ij} S_{ij}.
    \label{eq:wviscous-dissipation-term}
\end{equation}
Accordingly, the lubricant temperature evolution is governed by conduction 
from solid surfaces and convective transport in the fluid.
\subsection{Contact factor} \label{subsec:contact-factor} 

As mentioned in the introduction, a contact factor is introduced to quickly and intuitively
assess the extent of interaction between the gear and the lubricant during operation. 
The gear wetting process is conceptualized as progressive immersion, 
where the gear surface gradually comes into contact with the surrounding lubricant. 
To simplify this complex phenomenon, we estimate the instantaneous degree of contact 
using the following metric:

\begin{equation}
    \zeta_{C} = \frac{\sum_j V_j  W_{ij}}{\sum_{\substack{j \in \mathcal{N}_i^{\text{full}}}} V_j W_{ij}}
 \label{eq:Thermal-T-SPH-with-wall}
\end{equation}
where $j$ denotes the lubricant particles surrounding the gear particle $i$, 
$V_j$ is the volume of particle $j$, $W_{ij}$ is the value of the kernel function 
$W(\left |\mathbf{r}_{ij} \right |, h)$, which depends on the relative distance 
between particles$j$ and $i$. The denominator,
$\sum_{\substack{j \in \mathcal{N}_i^{\text{full}}}} V_j W_{ij}$ represents the case where 
the gear particle is fully immersed in lubricant.

This contact factor offers a macroscopic indicator of surface wetting and highlight 
regions experiencing reduced lubricant coverage. 
It is not intended to resolve micro-scale lubrication mechanisms or predict detailed 
film behavior; rather, it provides a simple, design-level metric that supports rapid 
assessment of lubricant distribution in complex transient conditions.

\subsection{Time integration} \label{subsec:time-integration} 
This work adopts the dual-criteria time-stepping approach \cite{method--zhang2020dual} to improve
computational efficiency. In this method, neighbor lists and kernel-related quantities are updated 
based on the advection time step, while the time integration of physical variables is constrained 
by the acoustic criterion.
The definitions of the advection time step $\Delta t_{ad}$ and the acoustic criterion $\Delta t_{ac}$
are as follows:
\begin{equation}
    \begin{cases}
   \Delta t_{ad}=CFL_{ad}min(\frac{h}{\lvert \mathbf{v} \rvert_{max}},\frac{h^2}{\nu}) \\
    \Delta t _{ac}=CFL_{ac}\frac{h}{c+\lvert \mathbf{v}_{max}\rvert}                              
    \end{cases},
    \label{eq:time-step-dual}
\end{equation}
where the $CFL_{ad} = 0.25$, $CFL_{ac}=0.6$, $\lvert v \rvert_{max}$ is the maximum particle
advection speed of the flow, and the $\nu$ is the kinematic viscosity of the fluid.

For the explicit integration of the thermal diffusion equation, 
the maximum allowable time step is given by:
\begin{equation}
    \Delta t_{T}=0.5(\frac{\rho C_p h^2}{k})
    \label{eq:time-step-temperature}
\end{equation}

%
%
\section{Validation of Numerical Methods with Experiments and Theoretic }
\label{sec:validation}
In this section, two validation cases are conducted to assess the accuracy and applicability 
of the proposed SPH-based multiphysics framework. 
The first validation compares the simulated churning losses with experimental data 
from a standard type C-PT FZG gearbox setup, 
demonstrating the capability of the method to accurately resolve transient 
oil–gear interactions under industrial conditions. 
The second validation examines the thermal solver using a flat-panel heat-conduction 
problem with an analytical solution, 
confirming its reliability in modeling heat transfer across materials with discontinuous properties.

\subsection{Validation with type C-PT FZG gearbox Data} \label{subsec:validation-FZG gear} 

\begin{figure}[htbp]
  \centering
  \begin{subfigure}[b]{0.48\textwidth}
    \includegraphics[width=0.9 \textwidth]{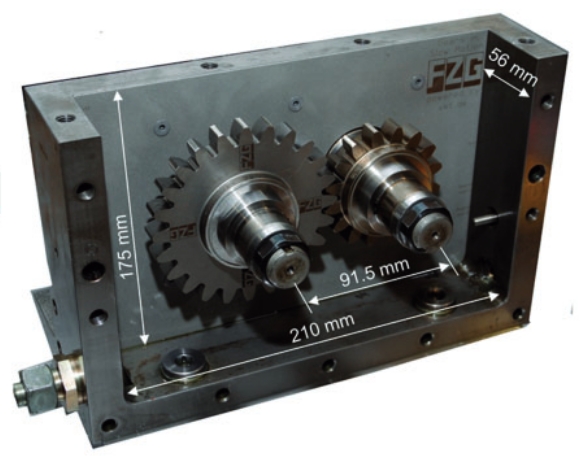}
    \caption{FZG gearbox used in experimental validation\cite{SPH--liu2019numerical}.}
    \label{fig:fzg-gearbox}
  \end{subfigure}
  \hfill
  \begin{subfigure}[b]{0.48\textwidth}
    \includegraphics[width=\textwidth]{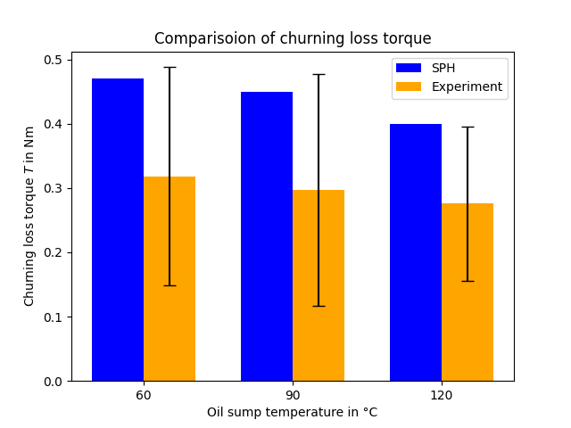}
    \caption{Comparison of churning loss torque from the SPH simulation and experiments.}
    \label{fig:churning-loss}
  \end{subfigure}
  \caption{Validation gearbox setup and simulation results from SPHinXsys compared with experimental data. 
  (a) Photograph of the experimental type C-PT FZG gearbox. 
  (b) Churning losses at various oil sump temperatures of simulation vs. experimental }
  \label{fig:fzg-gear-box-churning-loss}
\end{figure}

To validate the proposed SPH method under realistic operating conditions, 
oil splash simulations were conducted based on the Type C-PT FZG gearbox configuration. 
Figure \ref{fig:fzg-gearbox} presents a schematic of the C-PT FZG gearbox, 
while Table \ref{table:geometry-fzg-gear} lists the corresponding geometric parameters. 
Table \ref{table:operation-fzg-gear} summarizes the operating conditions used in both 
our simulations and the experimental study in Ref. \cite{SPH--liu2019numerical}.

The oil sump pressure was maintained at $1$ $bar$ across all cases, 
while the lubricant temperatures were set to \SI{60}{\celsius}, \SI{90}{\celsius}, 
and \SI{120}{\celsius}. 
The corresponding oil fill levels were $21.8 \mathrm{mm}$, $20.0 \mathrm{mm}$, 
and $18.1 \mathrm{mm}$ below the gearbox center axis, respectively. 
The rotational speed of the wheel was fixed at $1444 \mathrm{r/min}$ in all scenarios, 
resulting in a circumferential speed of $8.3 \mathrm{m/s}$ at the pitch circle.

Figure \ref{fig:churning-loss} compares the simulated churning loss torques 
obtained using SPHinXsys with experimental measurements. 
The predicted values (blue bars) show good agreement with the experimental 
data (orange bars), with all simulation results falling within the experimental 
uncertainty range. This consistency demonstrates that the proposed method can 
reliably capture lubricant–gear interaction and associated energy losses 
under thermally varying conditions.

\begin{table}[htbp]
  \centering
  \caption{Geometry parameters of the type C-PT FZG gearbox \cite{SPH--liu2019numerical}}
  \label{table:geometry-fzg-gear}
  \resizebox{\textwidth}{!}{%
    \begin{tabular}{lccccccc}
      \toprule
      & 
      \makecell{Centre \\ distance \\ $a$ ($\mathrm{mm}$)} & 
      \makecell{Number of \\ teeth $z$} & 
      \makecell{Normal \\ module \\ $m_n$ ($\mathrm{mm}$)} & 
      \makecell{Working \\ pressure \\ angle $\alpha$ (°)} & 
      \makecell{Addendum \\ modification \\ coefficient \\ $x_1$, $x_2$} & 
      \makecell{Face \\ width \\ $b$ ($\mathrm{mm}$)} & 
      \makecell{Tip \\ diameter \\ $d_a$ ($\mathrm{mm}$)} \\
      \midrule
      Pinion & 91.5 & 16 & 4.5 & 20.0 & 0.182 & 14 & 82.46 \\
      Wheel  &      & 24 & 4.5 &       & 0.171 & 14 & 118.36 \\
      \bottomrule
    \end{tabular}%
  }
\end{table}

\begin{table}[htbp]
  \centering
  \caption{Operating conditions of type C-PT FZG gearbox \cite{SPH--liu2019numerical}}
  \label{table:operation-fzg-gear}
  \resizebox{\textwidth}{!}{
    \begin{tabular}{ccccc}
      \toprule
      \makecell{Pressure \\ $p$ ($\mathrm{bar}$)} & 
      \makecell{Oil sump \\ temperature ($\mathrm{°C}$)} & 
      \makecell{Oil fill \\ level} & 
      \makecell{Rotational speed \\ of the wheel \\$n_2$ ($\mathrm{r/min}$)} & 
      \makecell{Circumferential \\ speed $v$ \\ at pitch circle ($\mathrm{m/s}$)} \\
      \midrule
      1 & 60  & \makecell{21.8 mm below \\ the middle axis} & 1444 & 8.3 \\
      1 & 90  & \makecell{20.0 mm below \\ the middle axis} & 1444 & 8.3 \\
      1 & 120 & \makecell{18.1 mm below \\ the middle axis} & 1444 & 8.3 \\
      \bottomrule
    \end{tabular}
  }
\end{table}

\subsection{Heat transfer in Slabs with different materials} 
\label{subsec:validation-heat} 

\begin{figure}[htbp]
  \centering
  \begin{subfigure}[b]{0.48\textwidth}
    \includegraphics[width=0.9 \textwidth]{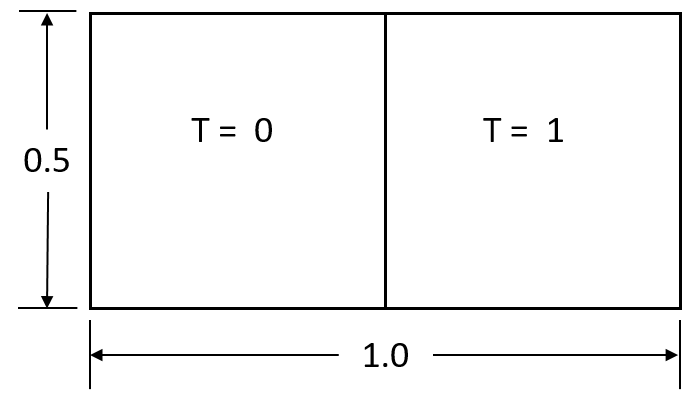}
    \caption{The initial setup of the slabs.}
    \label{fig:slabs}
  \end{subfigure}
  \hfill
  \begin{subfigure}[b]{0.48\textwidth}
    \includegraphics[width=\textwidth]{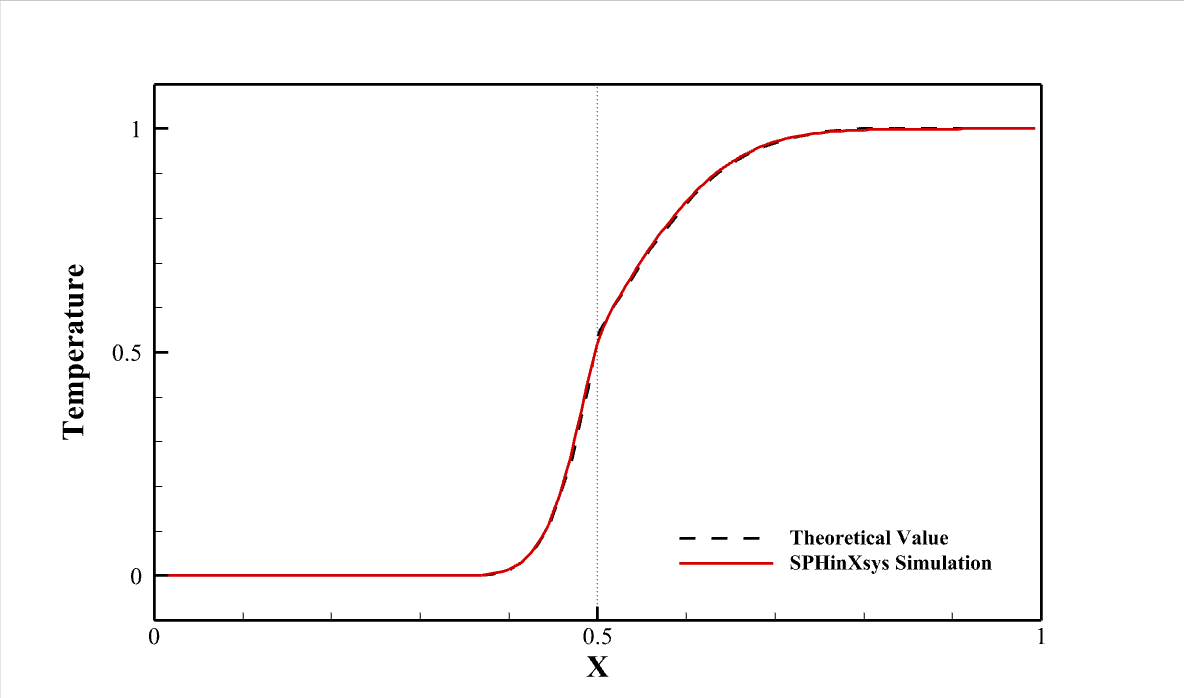}
    \caption{Temperature profile at t = 2.0}
    \label{fig:temperature-pl=2000-kr=3-t=2}
  \end{subfigure}
  \caption{Heat transfer in two slabs 
  (a) The geometry and initial temperature settings of the slabs. 
  (b) Temperature profile on the line of y = 0.25 at t = 2.0 }
  \label{fig:heat-transfer-two-slabs}
\end{figure}

To further assess the accuracy of the proposed SPH-based thermal solver, 
a classical transient heat conduction problem 
involving two materials with discontinuous thermal properties is simulated.
This benchmark case follows the configuration described in 
Refs.\cite{method--cleary1999conduction, validation--cleary1998heat--modelling}. 
As shown in Fig.\ref{fig:slabs}, 
the domain consists of a slab of unit length and half-unit width, 
with the material interface located at $x = 0.5$. 
Initially, the left half is assigned a dimensionless temperature of 0, and the right half to 1.

In this test, the two materials differ in density and thermal conductivity, 
while having identical specific heat capacities. The dimensionless properties 
used in the simulation are summarized in Table \ref{tab:thermal-properties-slab}, 
and the thermal diffusivities are calculated via $\alpha=\frac{k}{c_{p} \cdot \rho}$.
The particle spacing is set to $\Delta x = 0.0125$, and the smoothing length is defined as
$h= 1.0 \cdot \Delta x$.

\begin{table}[htbp]
  \centering
  \caption{Thermal properties of left and right materials}
  \begin{tabular}{ll}
    \hline
    Properties &  Values \\
    \hline
    Density of left material $\rho_l$ & 1000.0 \\
    Density of right material $\rho_r$ & 2000.0 \\
    Thermal conductivity of left material $k_l$ & 1.0 \\
    Thermal conductivity of right material $k_r$ & 3.0 \\
    Specific heat capacity of left material $c^l_p$ & 1.0 \\
    Specific heat capacity of right material $c^r_p$ & 1.0 \\
    Thermal diffusivity of left material $\alpha_l$ & 0.001 \\
    Thermal diffusivity of right material $\alpha_r$ & 0.0015 \\
    \hline
  \end{tabular}
  \label{tab:thermal-properties-slab}
\end{table}
The numerical results at time $t=2.0$ are shown in Fig.~\ref{fig:temperature-pl=2000-kr=3-t=2}, 
where the temperature profile along the centerline ($y=0.25$) is compared with the 
analytical solution. The SPH simulation accurately captures the temperature 
gradient and the discontinuity at the material interface, demonstrating the solver’s 
ability to handle material heterogeneity.

%
%
%
\section{Parametric Study and Performance Evaluation of the Bevel-Helical Gearbox Simulation}
\label{sec:Bevel-Helical-gearbox-simulation}
This section presents a comprehensive simulation study of a bevel-helical gearbox 
under various operating conditions, with the aim of analyzing lubricant dynamics, 
energy losses, thermal behavior, and computational efficiency.
First, the geometric configuration of the bevel-helical gearbox is introduced, 
followed by the specification of the variable combinations considered in the 
parametric study and a grid-independence analysis to determine an appropriate 
particle spacing for the subsequent simulations.
Second, the effects of various operating parameters on the performance of the bevel-helical
gearbox are systematically analyzed. Finally, the computational performance of the proposed
method is assessed by comparing the simulation efficiency on the GPU and the CPU platforms. 
\subsection{Geometry, Case Setup, and Grid Sensitivity Analysis}
\label{subsec:set-up-sensitivity}
\begin{figure}[htbp]
  \centering
  \begin{subfigure}[b]{0.48\textwidth}
    \includegraphics[width=\textwidth]{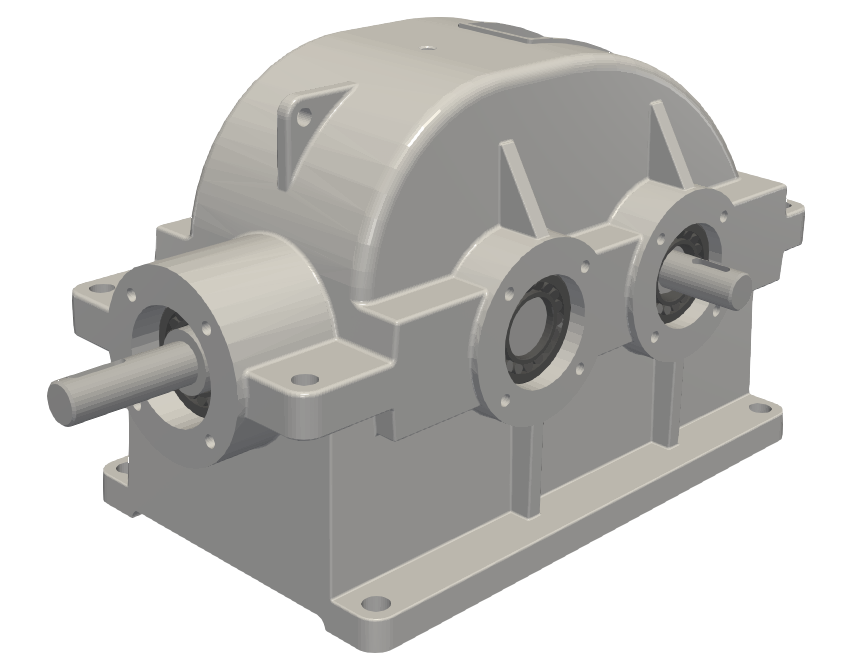}
    \caption{}
    \label{fig:gearbox-outlook}
  \end{subfigure}
  \hfill
  \begin{subfigure}[b]{0.48\textwidth}
    \includegraphics[width=\textwidth]{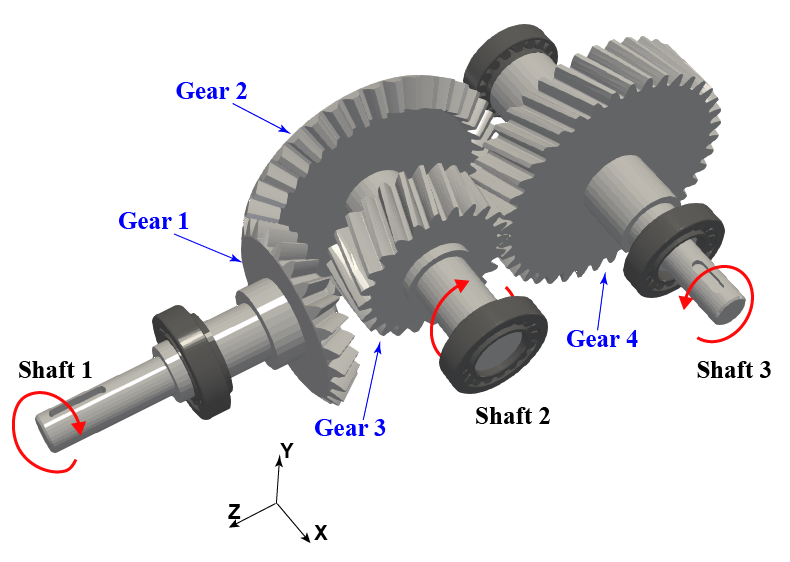}
    \caption{}
    \label{fig:three-shaft}
  \end{subfigure}
  \caption{Geometry and internal structure of the Bevel–Helical Gear Reducer. (a) External view of the gearbox housing. (b) Internal configuration showing the three-shaft system and gear pairs with rotational directions. }
  \label{fig:geometry-reducer}
\end{figure}

The target system in this study is a bevel–helical gear reducer featuring a three-shaft, 
two-stage configuration.
As shown in Figure \ref{fig:geometry-reducer}, the gearbox consists of a cast housing 
(Fig. \ref{fig:gearbox-outlook}) and an internal transmission system composed of two bevel 
gears and two helical gears mounted on Shaft 1 to Shaft 3 (Fig. \ref{fig:three-shaft}). 
Torque is transmitted from Shaft 1 to Shaft 3 via Shaft 2 through two gear stages. 
Shaft 1 serves as the input, while Shaft 3 is the output. 
The rotation directions of each shaft are indicated in Fig. \ref{fig:three-shaft} , 
and the gear pairs are meshed accordingly. 
For clarity, gear numbers are annotated in the figure to correspond with the parameter 
definitions in Table \ref{table:geometry-parameter-reducer}.
%

Table \ref{table:geometry-parameter-reducer} summarizes the geometric characteristics of 
the four gears, which define the meshing configuration and set the spatial constraints 
for the lubricant domain used in the simulation.

\begin{table}[htbp]
  \centering
  \caption{Geometry Parameters of the Bevel-Helical Gear Reducer }
  \label{table:geometry-parameter-reducer}
  \resizebox{\textwidth}{!}{
    \begin{tabular}{cccccccc}
      \toprule
      \makecell{Gear} & 
      \makecell{Type} & 
      \makecell{Number of \\ teeth $z$} & 
      \makecell{Tip Diameter \\ $d$ ($\mathrm{mm}$)} & 
      \makecell{Face width \\  $b$ ($\mathrm{mm}$)}& 
      \makecell{Helix angle \\  $\beta$ (°)}& 
      \makecell{Pitch cone  \\  angle $\delta$ (°)}& 
      \makecell{Normal module \\  $m_n$ ($\mathrm{mm}$)} \\
      \midrule
      Gear 1 & Bevel & 25 & 87.5 & 15 & {--} & 32 & 3.5 \\
      Gear 2 & Bevel & 40 & 140 & 15 & {--} & 58 & 3.5 \\
      Gear 3 & Helical & 24 & 82.6 & 30 & 18 & {--} & 3.5 \\
      Gear 3 & Helical & 40 & 133.7 & 30 & 18 & {--} & 3.5 \\
      \bottomrule
    \end{tabular}
  }
\end{table}

To systematically investigate the influence of operating parameters on lubricant behavior, 
churning losses, and thermal effect, a total of ten simulation cases are defined, 
as summarized in Table \ref{table:operating-conditions}. 
Cases 1 to 8 cover realistic industrial operating conditions with combinations 
of two shaft speeds (150 rad/s and 600 rad/s), 
two oil fill levels, defined as the vertical distances from the lubricant 
surface to the gear center, i.e., -0.036m and -0.05m,
and two kinematic viscosity values of the lubricant ($15 \mathrm{mm²/s}$ and $44 \mathrm{mm²/s}$).
The listed rotational speed in Table \ref{table:operating-conditions} refers to Shaft 3, 
and the reported maximum circumferential velocity corresponds to Gear 4.
Unless otherwise stated, all rotational speeds and tip velocities discussed hereafter 
follow the same conventions.
In addition, Case 9 and Case 10 are designed as the extreme conditions, 
where the shaft speed reaches 1500 rad/s and the resulting tip speed of the gear 4 
exceeds 100 m/s, matching the high-speed situation discussed in the introduction.
Case 10, on the other hand, represents a computationally intensive case under 
industrial-speed conditions. It employs a finer spatial resolution and a higher lubricant 
fill level, resulting in a significantly increased number of oil particles, on the order of 
several million, which imposes a heavier computational load.
The material properties required for the simulation are summarized in 
Table \ref{tab:thermal-properties}.

\begin{table}[htbp]
  \centering
 
  \caption{Operating conditions for gearbox lubrication simulation.}
  \label{table:operating-conditions}
  \resizebox{\textwidth}{!}{
    \begin{tabular}{cccccc}
      \toprule
      \makecell{Case} & 
      \makecell{Rotational speed \\ $\omega$ ($\mathrm{rad/s}$) / ($\mathrm{rpm}$) \textsuperscript{a} } & 
      \makecell{Max circumferential \\ velocity $v$ ($\mathrm{m/s}$)\textsuperscript{a} }& 
      \makecell{Lubrication depth $d$ ($\mathrm{m}$)\\ (distance from gear center)} & 
      \makecell{Kinematic viscosity \\ $\nu$ ($\mathrm{mm\textsuperscript{2}/s}$)} &
      \makecell{Resolution  \\ $d_p$ ($\mathrm{m}$)} \\
      \midrule
      Case 1 & 150 / 1432.4 & 10.05 & -0.05 & 15 & 0.0012\\
      Case 2 & 150 / 1432.4 & 10.05 & -0.05 & 44 & 0.0012\\
      Case 3 & 150 / 1432.4 & 10.05 & -0.036 & 15 & 0.0012\\
      Case 4 & 150 / 1432.4 & 10.05 & -0.036 & 44 & 0.0012\\
      Case 5 & 600 / 5729.6 & 40.10 & -0.05 & 15 & 0.0012\\
      Case 6 & 600 / 5729.6 & 40.10 & -0.05 & 44 & 0.0012\\
      Case 7 & 600 / 5729.6 & 40.10 & -0.036 & 15 & 0.0012\\
      Case 8 & 600 / 5729.6 & 40.10 & -0.036 & 44 & 0.0012\\
      Case 9 & 1500 / 14324 & 100.50 & -0.05 & 44 & 0.0012\\
      Case 10 & 600 / 5729.6 & 40.10 & -0.01 & 44 & 0.0008\\
      \bottomrule
    \end{tabular}
  }
  \vspace{0.5ex}
\begin{minipage}{\textwidth}
\scriptsize
\textsuperscript{a} Measured at Shaft 3. \\
\textsuperscript{b} Measured at the tip of Gear 4.
\end{minipage}
\end{table}

\begin{table}[htbp]
  \centering
  \small 
  \caption{Physical properties of lubricant and gears}
  \label{tab:thermal-properties}
  
    \begin{tabular}{lc}
      \toprule
      \makecell{Property} & \makecell{Value} \\
      \midrule
      Lubricant density $\rho$ ($\mathrm{kg/m^3}$) & 968.1 \\
      Gear shaft density $\rho$ ($\mathrm{kg/m^3}$) & 7850 \\
      Lubricant thermal conductivity $k$ ($\mathrm{W/(m\cdot K)}$) & 0.15 \\
      Gear shaft thermal conductivity $k$ ($\mathrm{W/(m\cdot K)}$) & 20 \\
      Lubricant specific heat $C_p$ ($\mathrm{J/(kg\cdot K)}$) & 2000 \\
      Initial temperature of lubricant $T_l$ ($\mathrm{K}$) & 298.15 \\
      Initial temperature of gear shaft $T_g$ ($\mathrm{K}$) & 318.15 \\
      \bottomrule
    \end{tabular}
  
\end{table}

\begin{figure}[htbp]
  \centering
  \begin{subfigure}[b]{0.48\textwidth}
    \includegraphics[width=\textwidth]{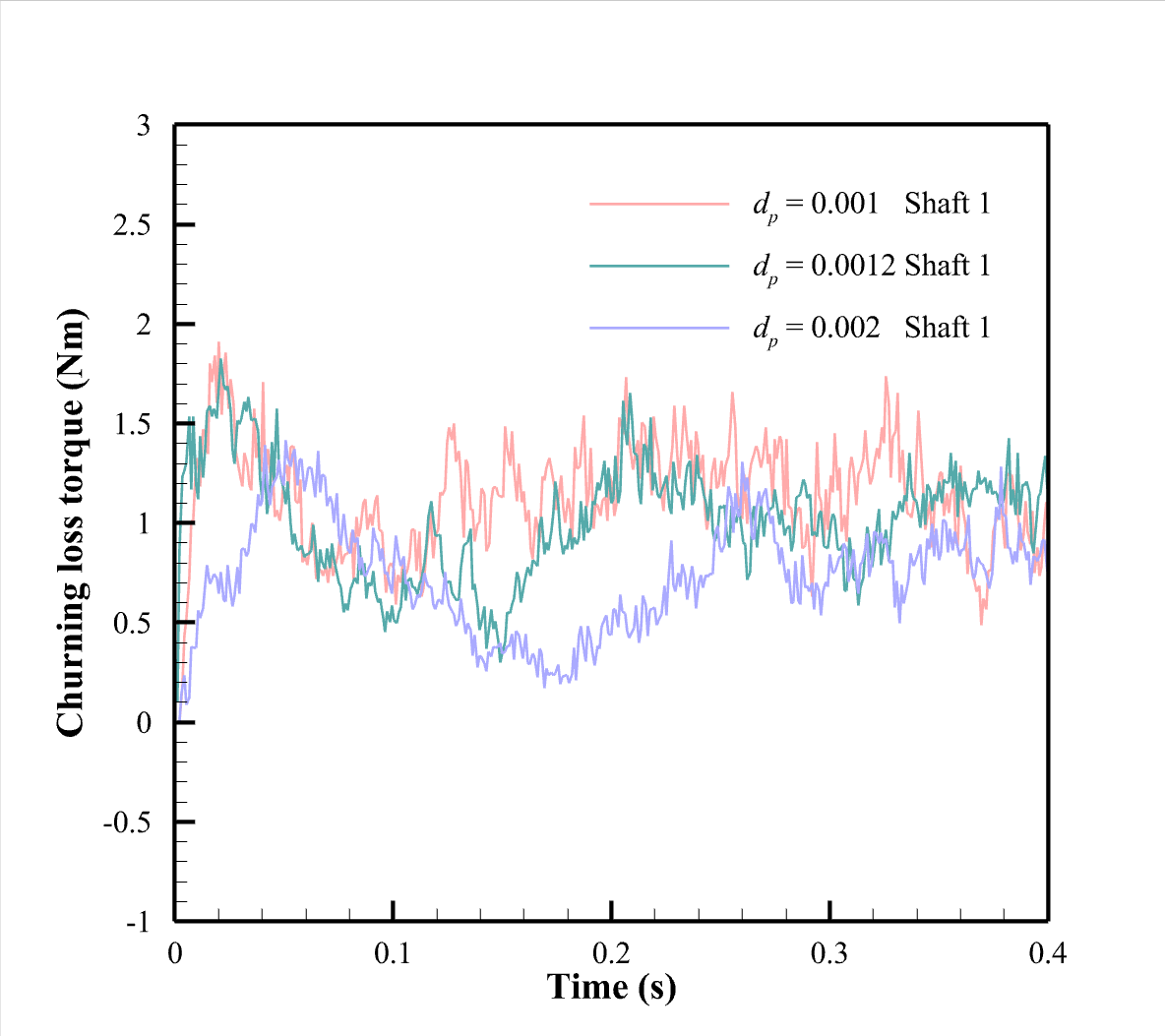}
    \caption{Churning loss torque of Shaft 1 over time for different particle spacings $d_p$}
    \label{fig:mesh-less-chruning-loss-shaft1}
  \end{subfigure}
  \hfill
  \begin{subfigure}[b]{0.48\textwidth}
    \includegraphics[width=\textwidth]{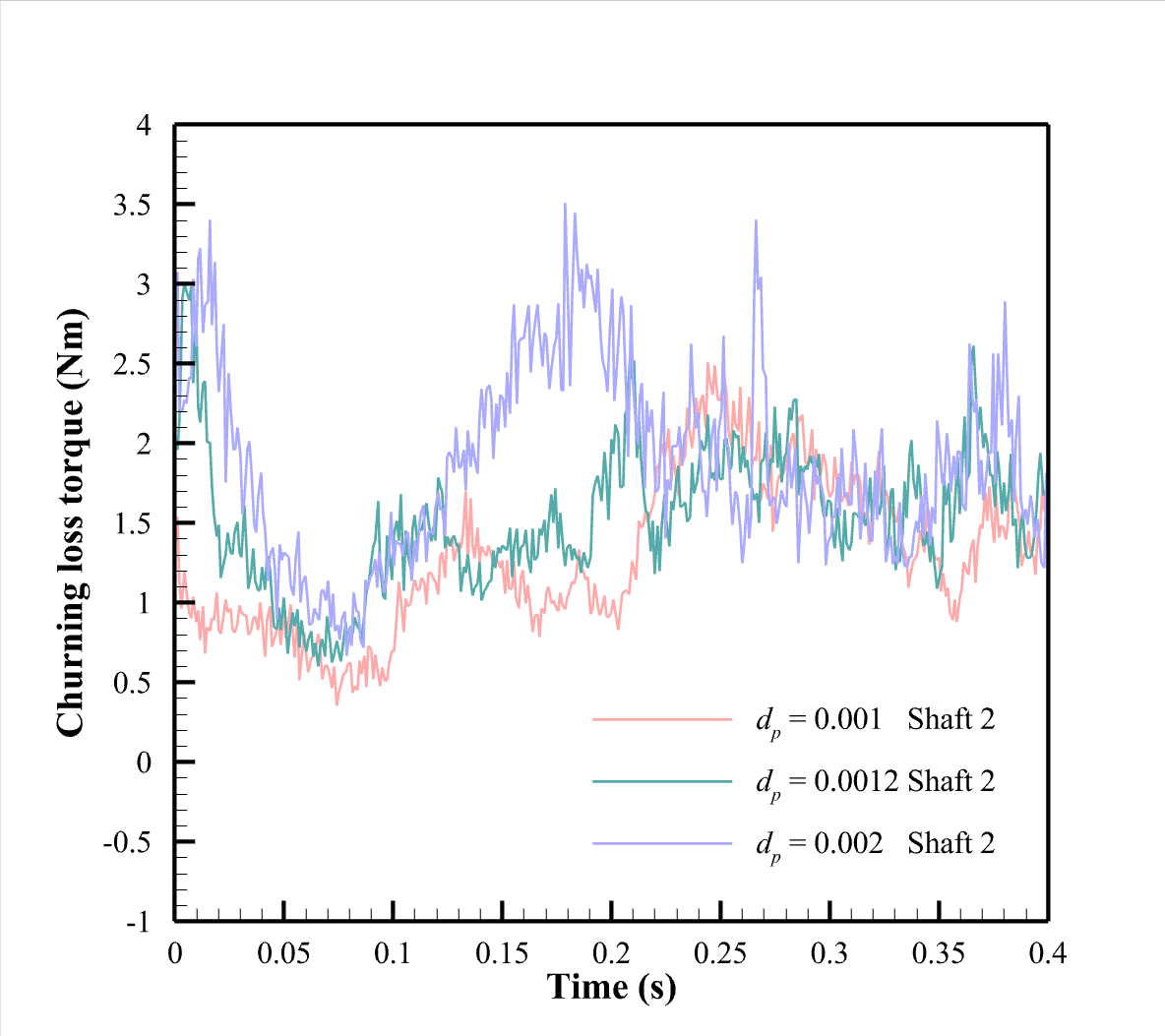}
    \caption{Churning loss torque of Shaft 2 over time for different particle spacings $d_p$}
    \label{fig:mesh-less-chruning-loss-shaft2}
  \end{subfigure}
  \vspace{0.3cm} 
  
  \begin{subfigure}[b]{0.48\textwidth}
    \includegraphics[width=\textwidth]{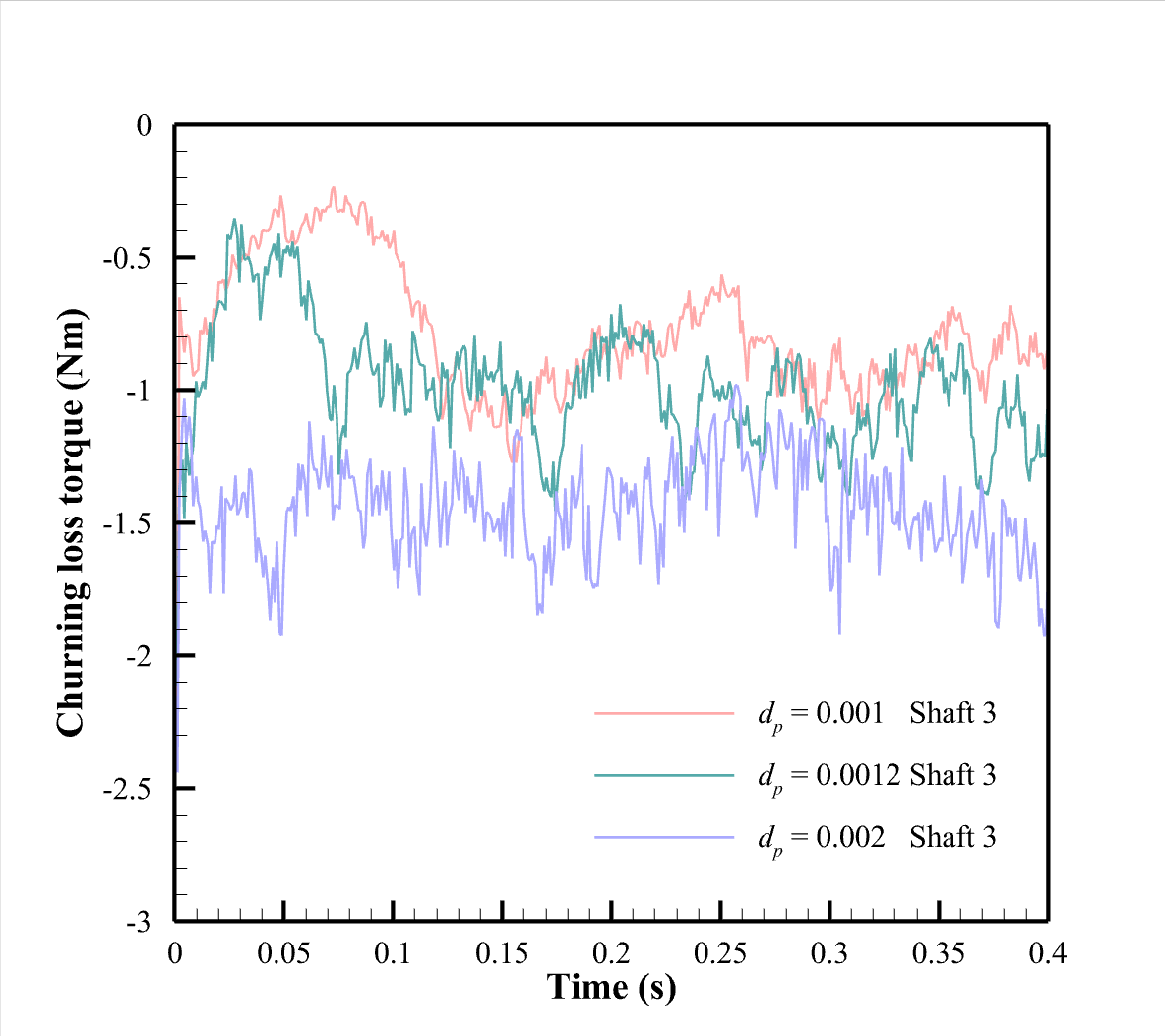}
    \caption{Churning loss torque of Shaft 3 over time for different particle spacings $d_p$}
    \label{fig:mesh-less-chruning-loss-shaft3}
  \end{subfigure}
  \hfill
  \begin{subfigure}[b]{0.48\textwidth}
    \includegraphics[width=\textwidth]{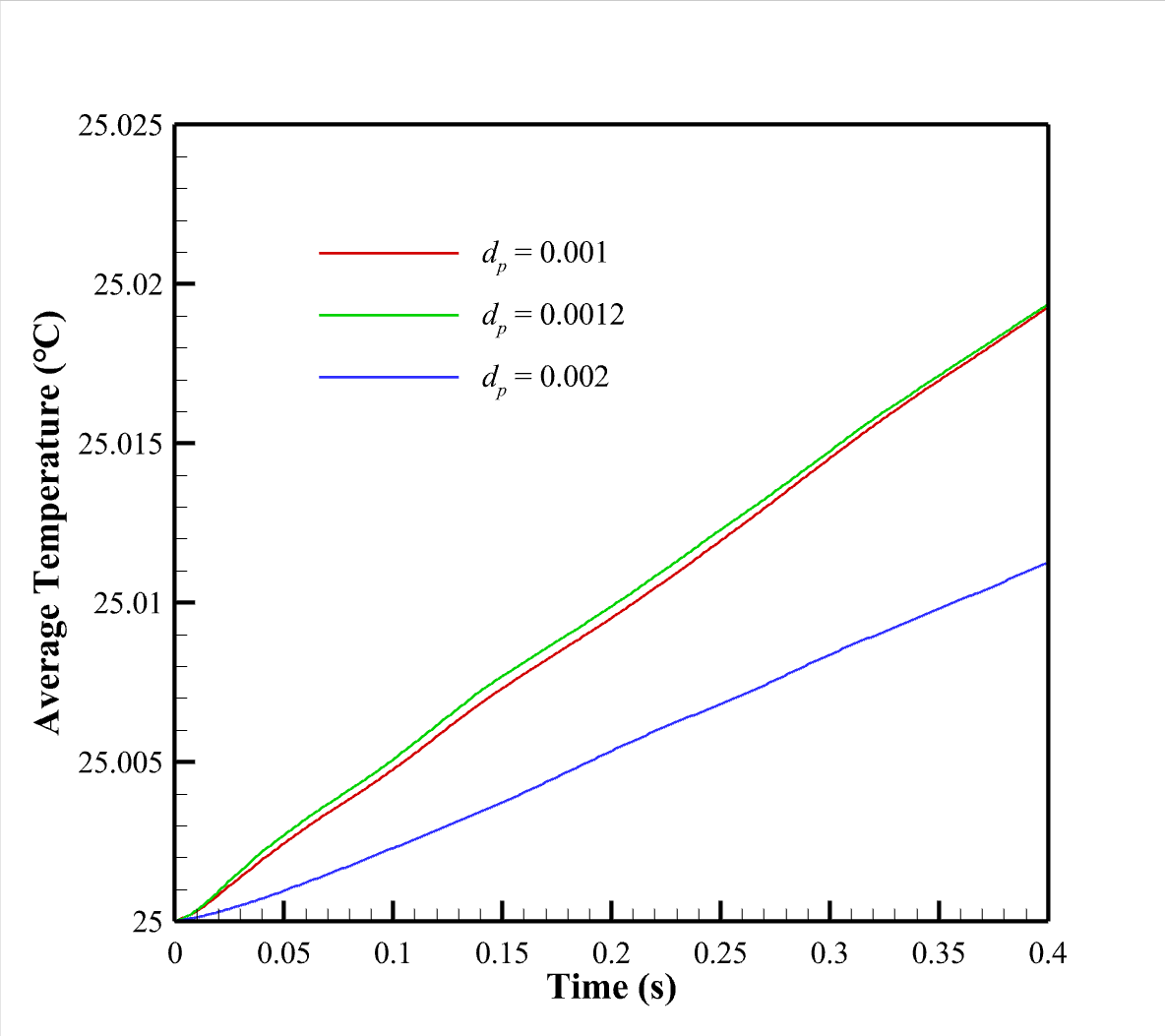}
    \caption{Evolution of average lubricant temperature for different particle spacings $d_p$}
    \label{fig:mesh-less-lubricant-temperature}
  \end{subfigure}
  \caption{Mesh independence verification for churning loss torque and lubricant temperature arising. 
  (a – c) Torque histories for Shafts 1–3 at particle spacings $d_p$ = 0.001, 0.0012, and 0.002 m.
  (d) Average lubricant temperature for the same resolutions.
}
  \label{fig:gearbox-meshless-validation}
\end{figure}

Before conducting the parametric studies, a grid independence analysis is carried 
out to determine a suitable particle spacing $d_p$. 
This mesh independence validation case adopts the same operating conditions as 
Case 1 in Table \ref{table:operating-conditions}.
Three particle spacings (0.001$\mathrm{m}$, 0.0012$\mathrm{m}$, and 0.002$\mathrm{m}$) are evaluated, 
corresponding to lubricant particle counts on the order of $10^6$, $5 \times 10^5$,
and $10^5$, respectively.
These resolutions are chosen to span a representative range of computational 
loads while capturing key physical features.
Figures \ref{fig:mesh-less-chruning-loss-shaft1} - \ref{fig:mesh-less-chruning-loss-shaft3} 
present the time histories of churning loss torque on Shaft 1 - Shaft 3. 
Although all resolutions exhibit strong transient fluctuations inherent to the oil–gear interaction, 
the torque responses show clear grid-convergence trends in terms of their large-scale temporal behavior. 
As the particle spacing decreases, the mean torque level and the overall waveform shape become increasingly similar. 
The curves for $d_p =0.001 \mathrm{m}$ and $d_p =0.0012 \mathrm{m}$ nearly coincide, 
whereas $d_p =0.002 \mathrm{m}$ shows a larger deviation but retains the same qualitative characteristics.

This convergence is more clearly reflected in the average lubricant temperature evolution 
shown in Figure \ref{fig:mesh-less-lubricant-temperature}, where the results from $d_p = 0.001 \mathrm{m}$
and $d_p =0.0012 \mathrm{m}$ are nearly indistinguishable over the simulated period. 
In view of this, the particle spacing $d_p =0.0012 \mathrm{m}$ is adopted in all subsequent 
simulations as a balanced choice between computational cost and numerical fidelity.

\subsection{Parametric Investigation of Oil–Gear Interaction in Industrial Conditions}
\label{subsec:Parametric-Investigation-Industrial-Conditions}
To assess the influence of various operating parameters on gearbox performance, 
this section analyzes simulation results from Case 1 to Case 8 in Table \ref{table:operating-conditions}, 
which represent typical industrial working conditions. 
The focus is on evaluating the churning losses on each shaft, the dynamic behavior of lubricant splashing, 
the evolution of oil–gear contact, and the lubricant temperature arising effect.

\begin{figure}[htbp]
  \centering
  \begin{subfigure}[b]{0.49\textwidth}
    \includegraphics[width=\textwidth]{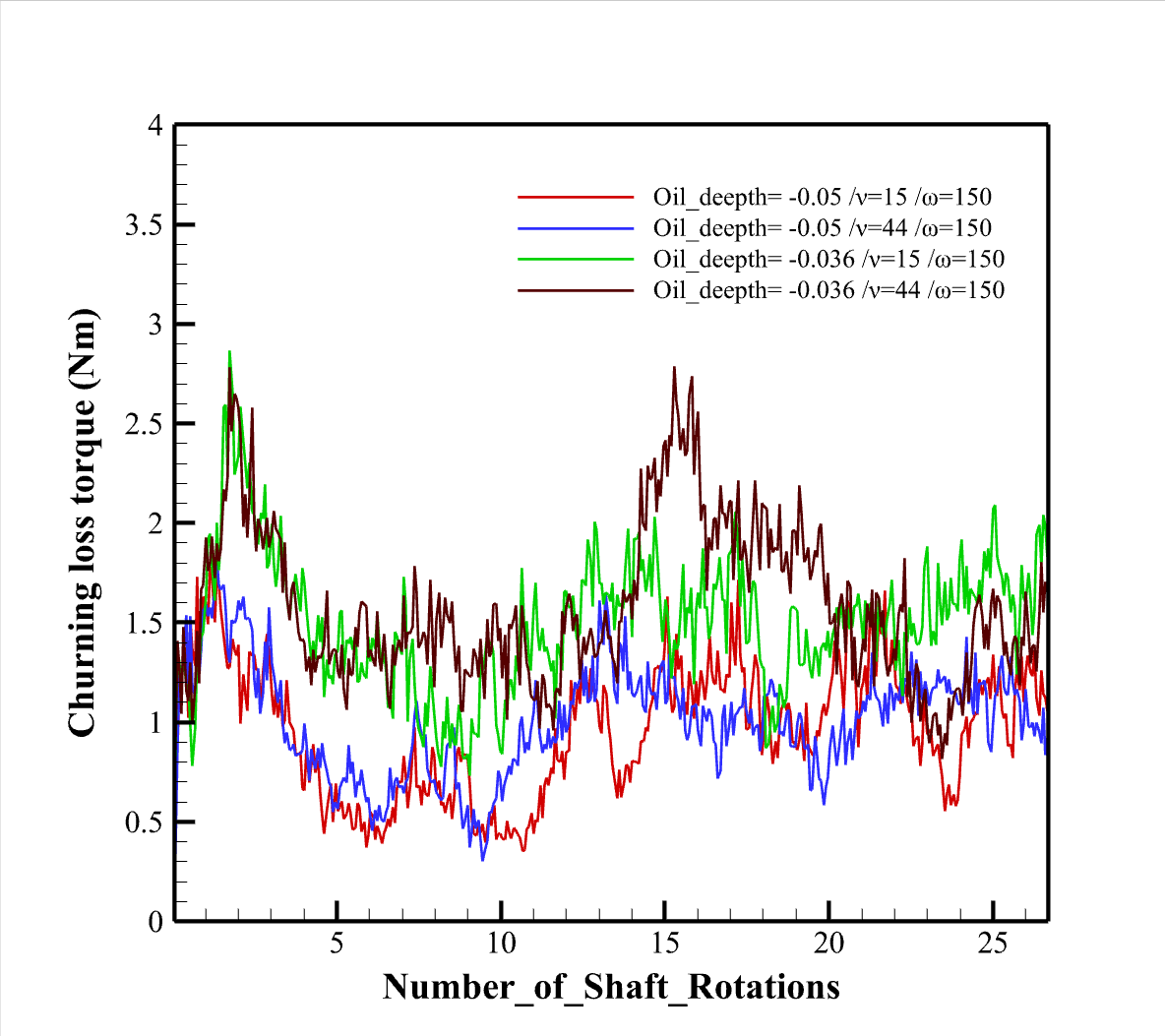}
    \caption{Churning loss on Shaft 1 while $\omega_3 = 150$ $\mathrm{rad/s}$}
    \label{fig:churning-loss-shaft1-150rad}
  \end{subfigure}
  \hfill
  \begin{subfigure}[b]{0.49\textwidth}
    \includegraphics[width=\textwidth]{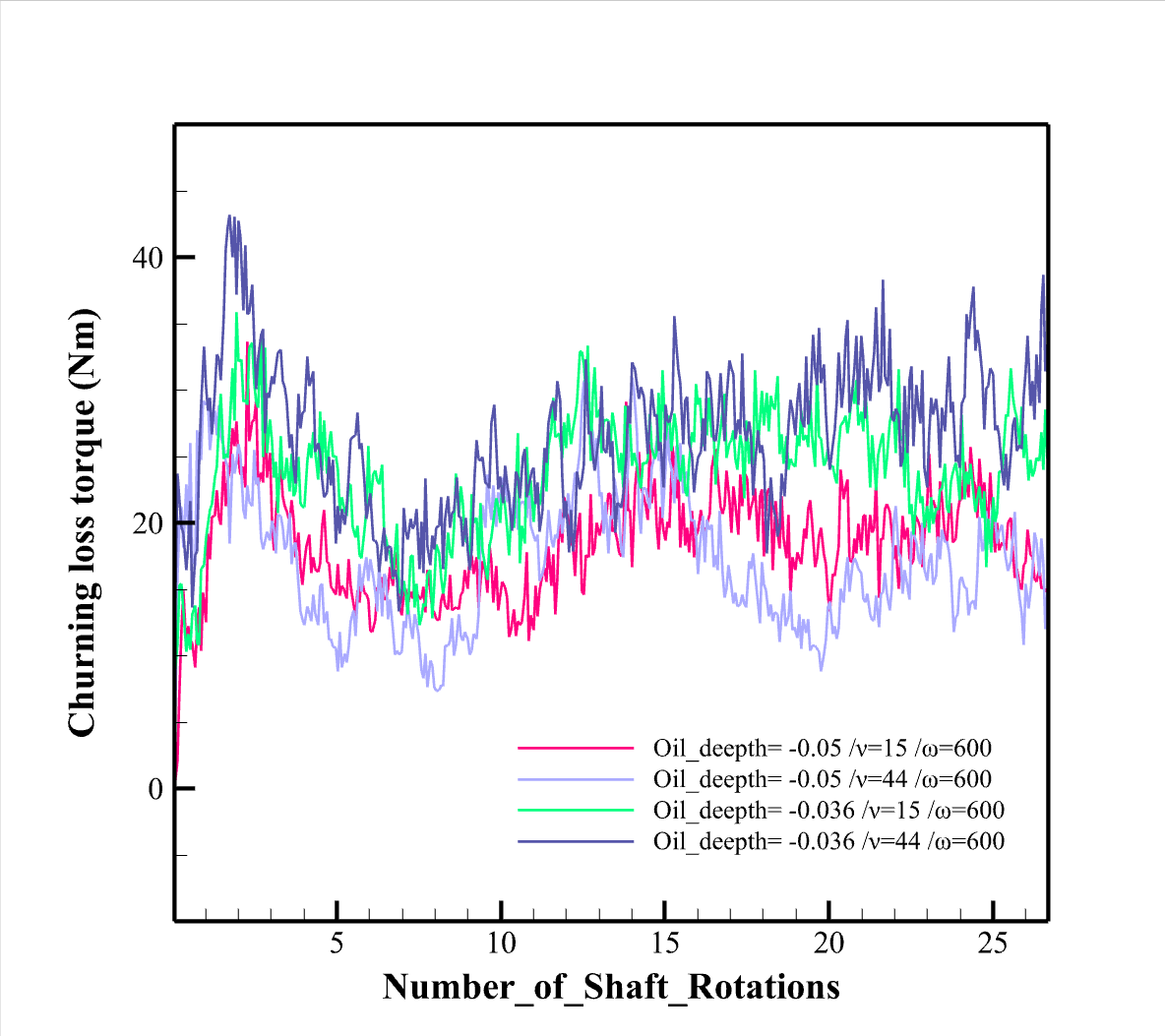}
    \caption{Churning loss on Shaft 1 while $\omega_3 = 600$ $\mathrm{rad/s}$}
    \label{fig:churning-loss-shaft1-600rad}
  \end{subfigure}

  \vspace{0.3cm}
  
  \begin{subfigure}[b]{0.49\textwidth}
    \includegraphics[width=\textwidth]{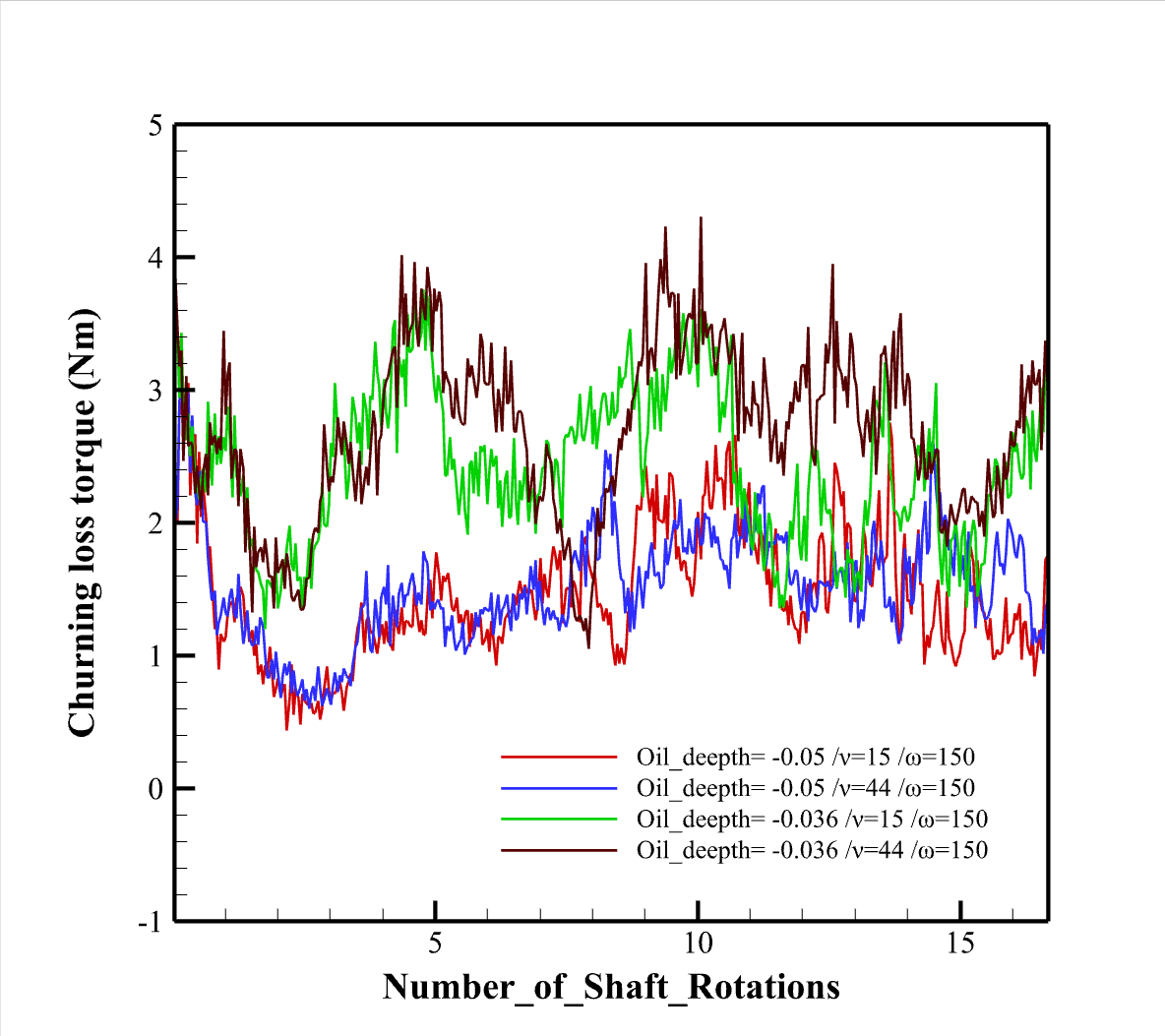}
    \caption{Churning loss on Shaft 2 while $\omega_3 = 150$ $\mathrm{rad/s}$}
    \label{fig:churning-loss-shaft2-150rad}
  \end{subfigure}
  \hfill
  \begin{subfigure}[b]{0.49\textwidth}
    \includegraphics[width=\textwidth]{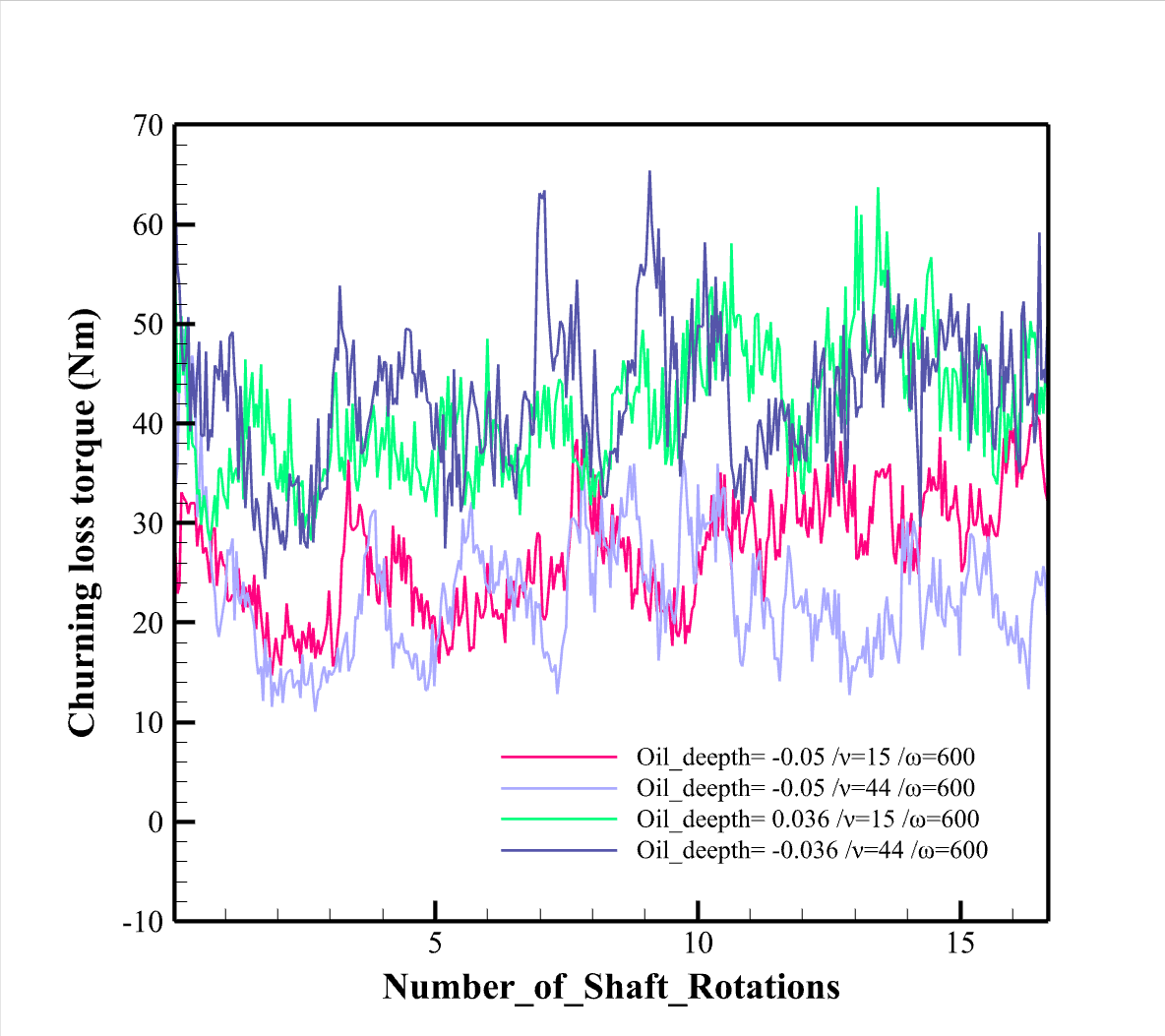}
    \caption{Churning loss on Shaft 2 while $\omega_3 = 600$ $\mathrm{rad/s}$}
    \label{fig:churning-loss-shaft2-600rad}
  \end{subfigure}

  \vspace{0.3cm}

  \begin{subfigure}[b]{0.49\textwidth}
    \includegraphics[width=\textwidth]{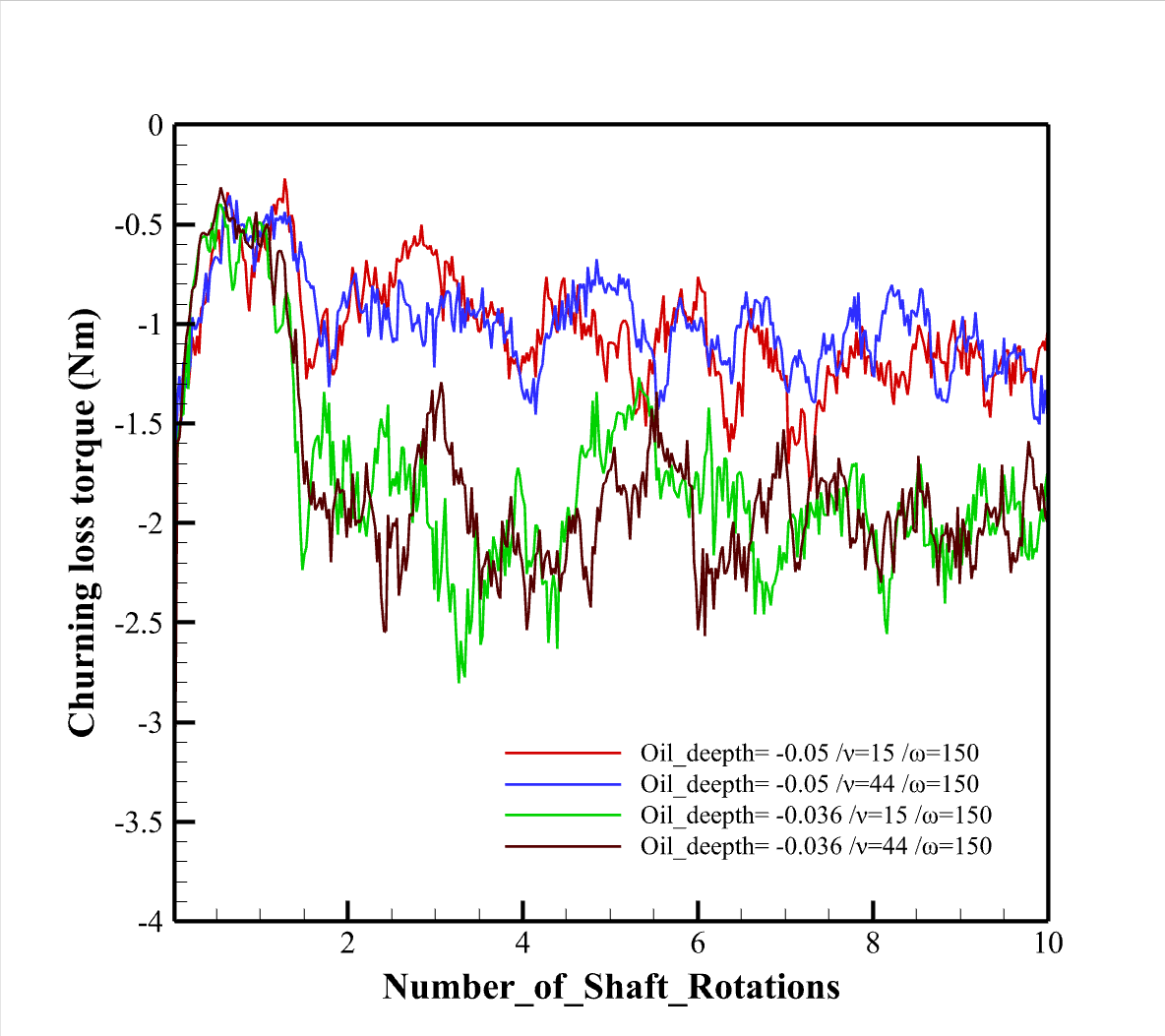}
    \caption{Churning loss on Shaft 3 while $\omega_3 = 150$ $\mathrm{rad/s}$}
    \label{fig:churning-loss-shaft3-150rad}
  \end{subfigure}
  \hfill
  \begin{subfigure}[b]{0.49\textwidth}
    \includegraphics[width=\textwidth]{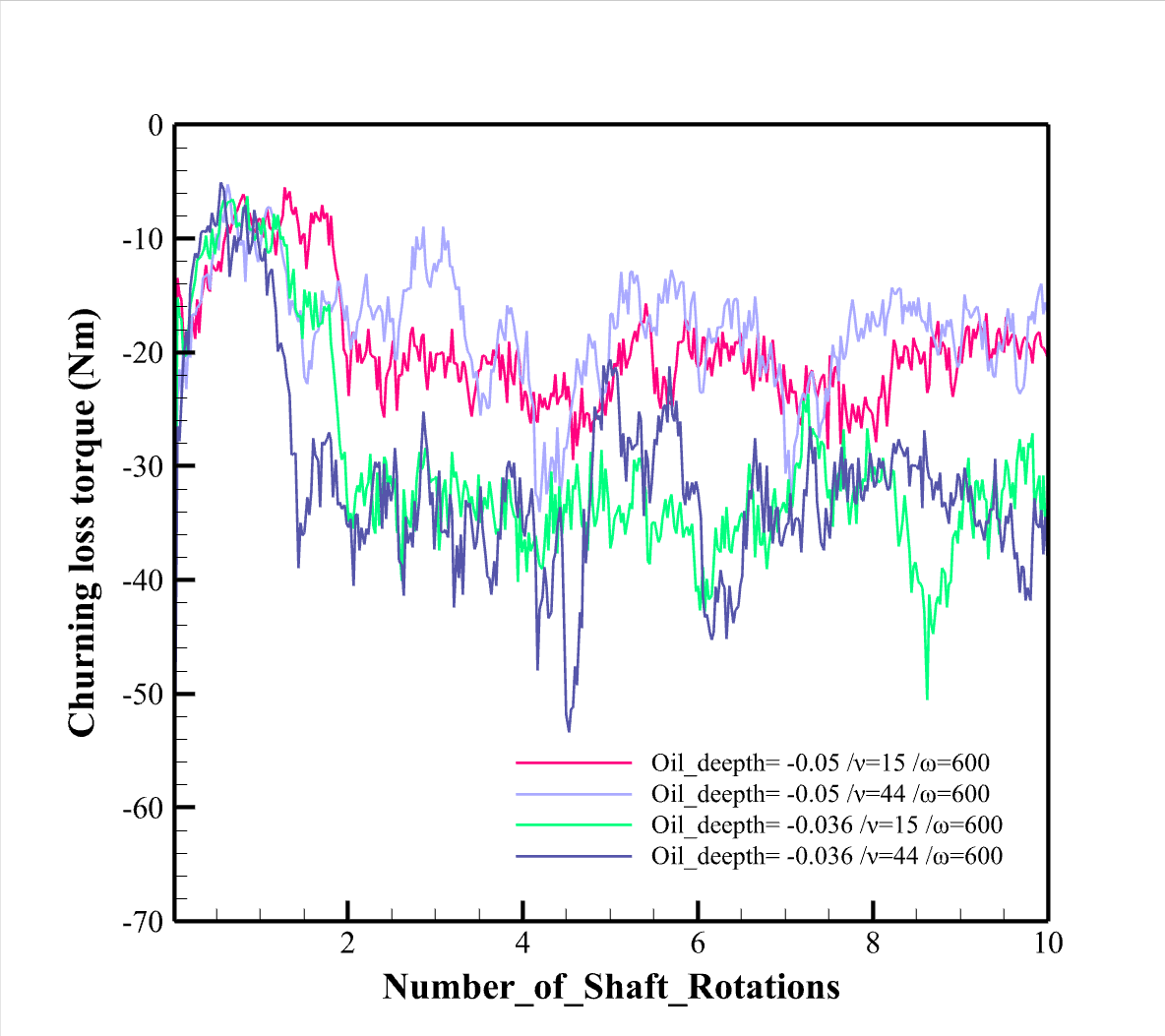}
    \caption{Churning loss on Shaft 3 while $\omega_3 = 600$ $\mathrm{rad/s}$}
    \label{fig:churning-loss-shaft3-600rad}
  \end{subfigure}

  \caption{Churning-loss torque on Shafts 1–3 under two operating speeds. 
  (a-c) Torque histories for Shaft 1-3 when Shaft 3 rotates at $\omega = 150$ $\mathrm{rad/s}$.
  (b-f) Torque histories for Shaft 1-3 when Shaft 3 rotates at $\omega = 600$ $\mathrm{rad/s}$. 
  The horizontal axis is normalized by the number of revolutions completed by each shaft
  (10 revolutions of Shaft 3 correspond to 26.67 of Shaft 1 and 16.67 of Shaft 2).}
  \label{fig:churning-loss-all}
\end{figure}

Figure \ref{fig:churning-loss-all} presents the time histories of churning loss torque on Shaft 1, 
Shaft 2, and Shaft 3 across all eight cases. Given the significant differences in torque magnitude 
at different rotational speeds, those curves are grouped and plotted separately for $\omega = 150$ $\mathrm{rad/s}$ 
and $\omega = 600$ $\mathrm{rad/s}$, respectively. It is important to note that all simulations are executed 
until Shaft 3 completes exactly 10 full revolutions, corresponding to 26.67 revolutions of Shaft 1 
and 16.67 revolutions of Shaft 2. To ensure consistent comparison across different angular speeds, 
the horizontal axis is normalized by the number of revolutions for each 
respective shaft in all figures.

As observed in Figure \ref{fig:churning-loss-all}, the torque histories generally stabilize after 
approximately two revolutions of Shaft 3, after which they fluctuate around quasi-steady levels. 
In this regime, several tendencies can be identified. 
First, varying the lubricant viscosity between $15$ and $44$ $\mathrm{mm²/s}$ 
only weakly affects the mean churning torque compared with the influence of shaft speed and oil fill level. 
Second, cases with a higher oil fill level systematically exhibit larger torque on all shafts, 
indicating increased hydrodynamic resistance due to the larger volume of oil being entrained. 
Third, for otherwise identical conditions, increasing the shaft speed from $150$ $\mathrm{rad/s}$ 
to $600$ $\mathrm{rad/s}$ leads to a markedly higher churning loss on each shaft, i.e., 
the torque levels at $600$ $\mathrm{rad/s}$ are several times 
those at $150$ $\mathrm{rad/s}$. These tendencies highlight the need to balance oil quantity and viscosity when 
optimizing gearbox efficiency under high-speed operation.

\begin{figure}[htbp]
  \centering
  \begin{subfigure}[b]{0.48\textwidth}
    \includegraphics[width=\textwidth]{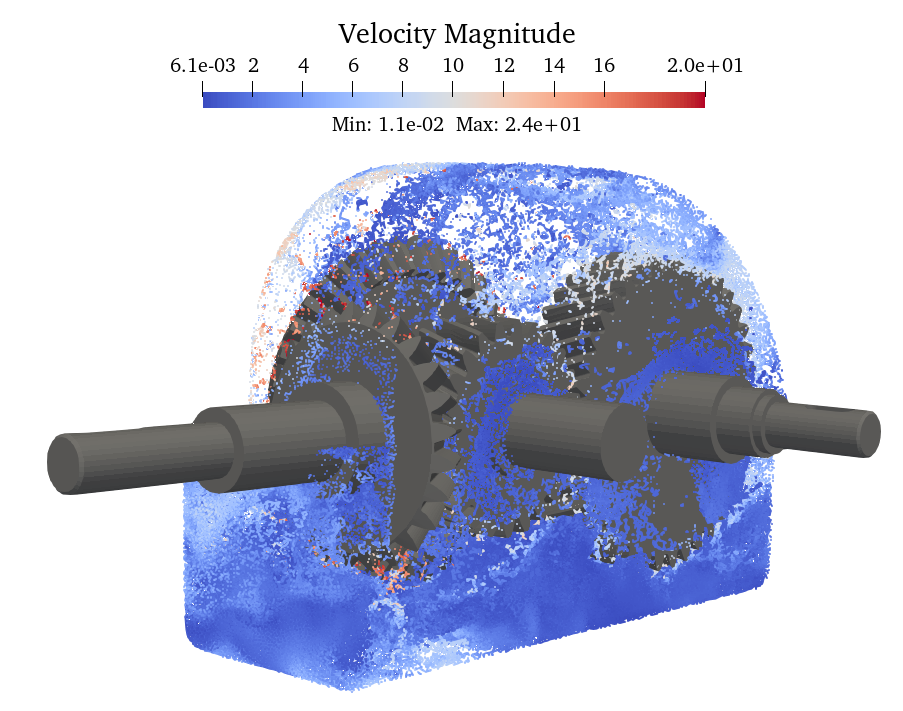}
    \caption{Oil-depth = -0.05 \textbackslash $\nu$ = 15 \textbackslash $\omega$ = 150 }
    \label{fig:velocity-Oil-depth-0.05-15-150}
  \end{subfigure}
  \hfill
  \begin{subfigure}[b]{0.48\textwidth}
    \includegraphics[width=\textwidth]{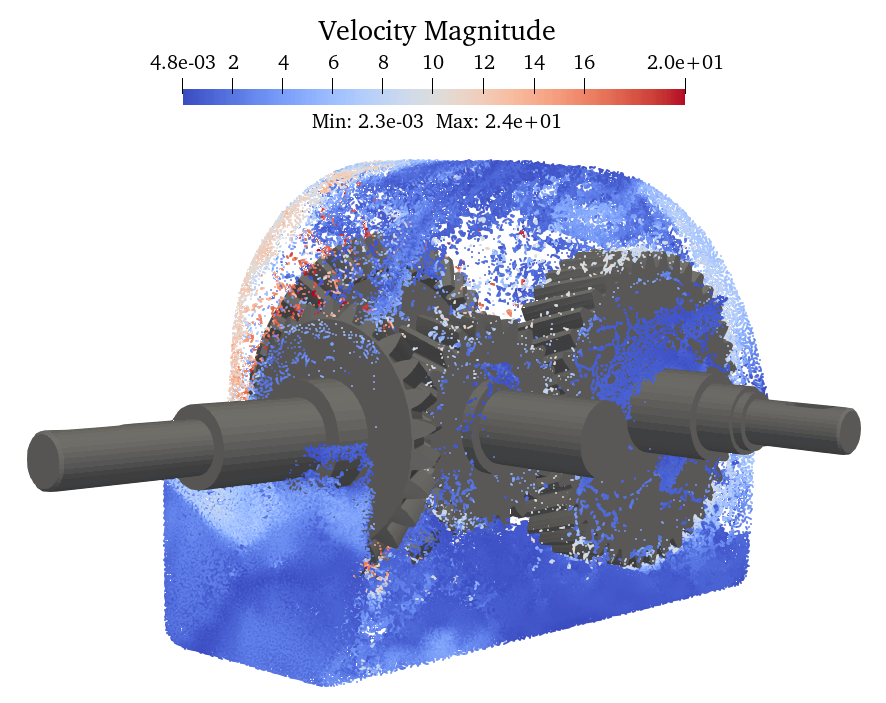}
    \caption{Oil-depth = -0.05 \textbackslash $\nu$ = 44 \textbackslash $\omega$ = 150}
    \label{fig:velocity-Oil-depth-0.05-44-150}
  \end{subfigure}
  \vspace{0.3cm} 
  
  \begin{subfigure}[b]{0.48\textwidth}
    \includegraphics[width=\textwidth]{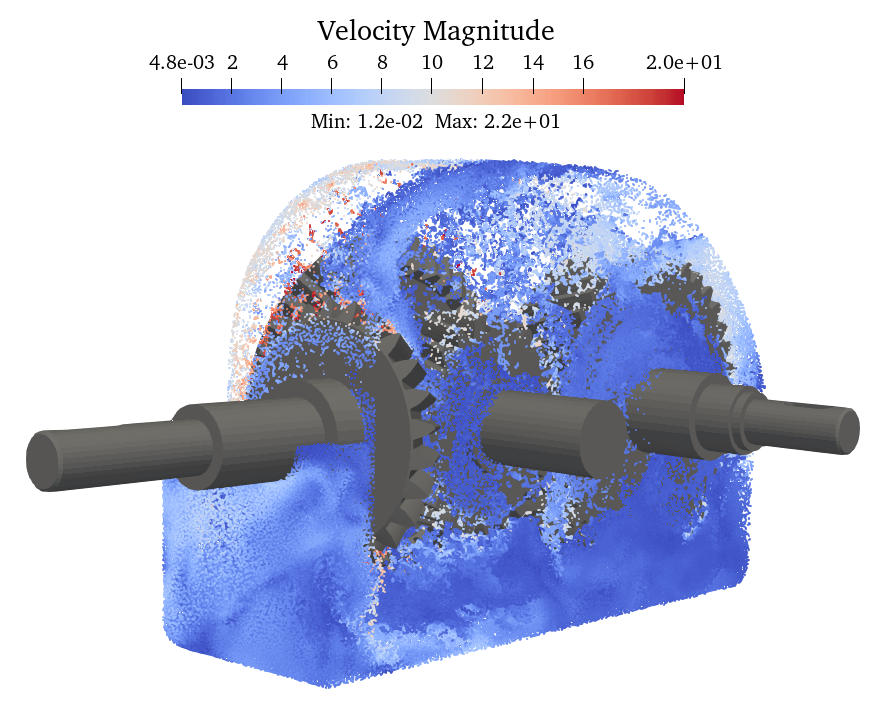}
    \caption{Oil-depth = -0.036 \textbackslash $\nu$ = 15 \textbackslash $\omega$ = 150}
    \label{fig:velocity-Oil-depth-0.036-15-150}
  \end{subfigure}
  \hfill
  \begin{subfigure}[b]{0.48\textwidth}
    \includegraphics[width=\textwidth]{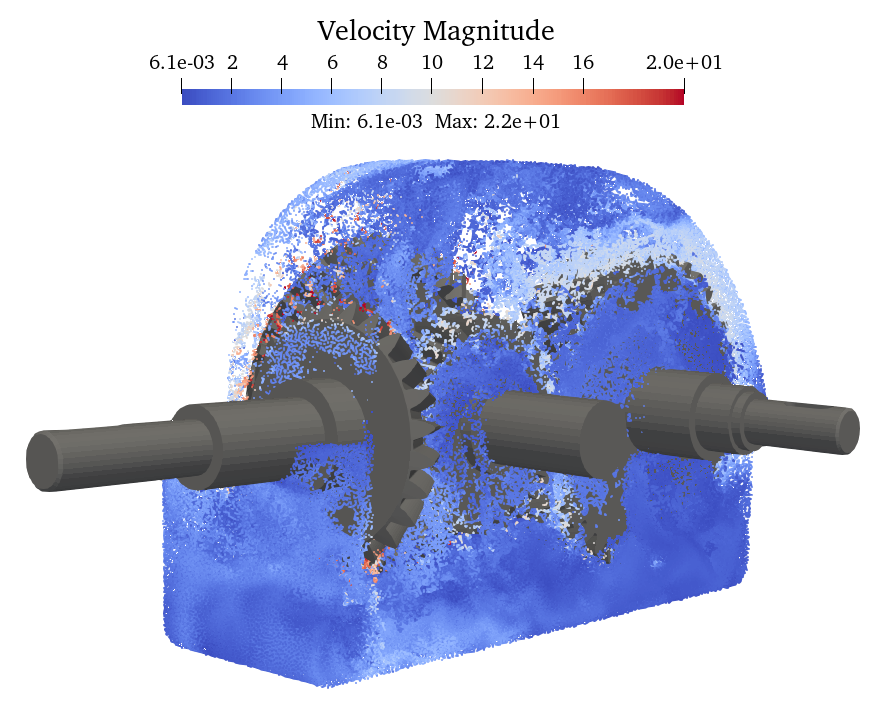}
    \caption{Oil-depth = -0.036 \textbackslash $\nu$ = 44 \textbackslash $\omega$ = 150}
    \label{fig:velocity-Oil-depth-0.036-44-150}
  \end{subfigure}

  \begin{subfigure}[b]{0.48\textwidth}
    \includegraphics[width=\textwidth]{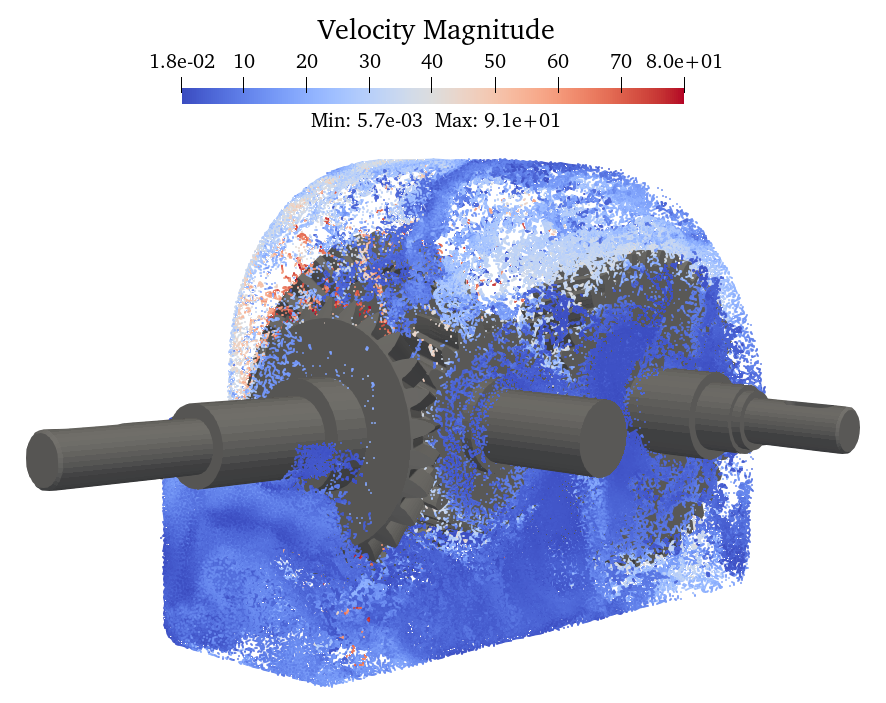}
    \caption{Oil-depth = -0.05 \textbackslash $\nu$ = 15 \textbackslash $\omega$ = 600}
    \label{fig:velocity-Oil-depth-0.05-15-600}
  \end{subfigure}
  \hfill
  \begin{subfigure}[b]{0.48\textwidth}
    \includegraphics[width=\textwidth]{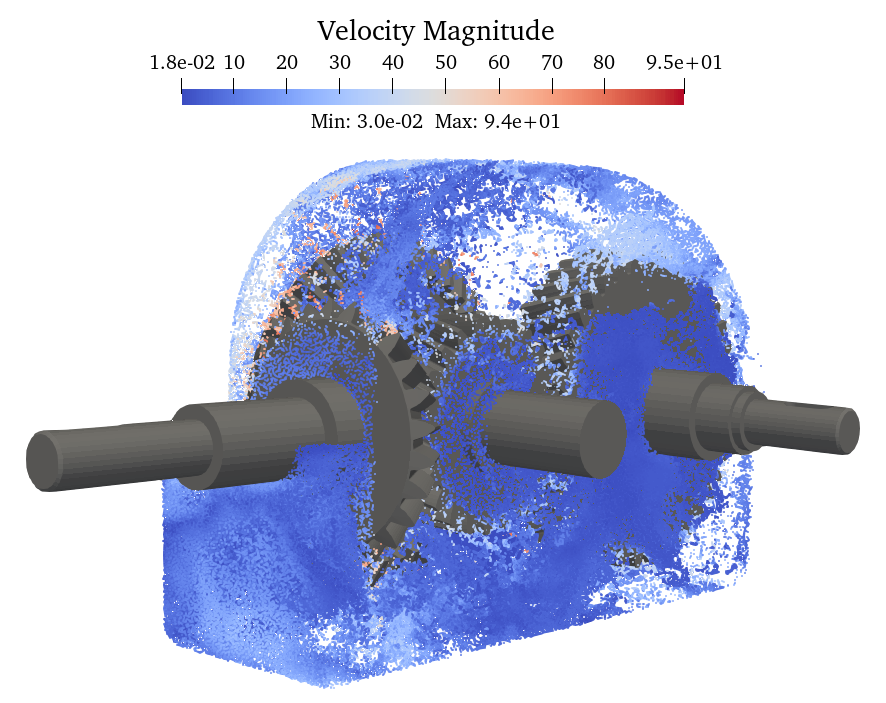}
    \caption{Oil-depth = -0.05 \textbackslash $\nu$ = 44 \textbackslash $\omega$ = 600}
    \label{fig:velocity-Oil-depth-0.0-44-600}
  \end{subfigure}
  \vspace{0.3cm} 
  
  \begin{subfigure}[b]{0.48\textwidth}
    \includegraphics[width=\textwidth]{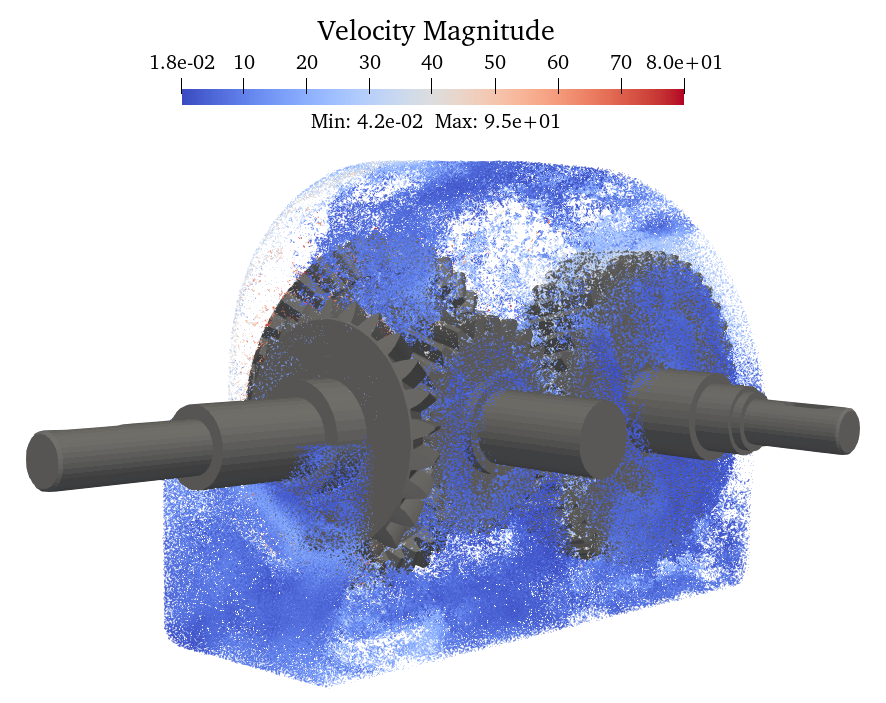}
    \caption{Oil-depth = -0.036 \textbackslash $\nu$ = 15 \textbackslash $\omega$ = 600}
    \label{fig:velocity-Oil-depth-0.036-15-600}
  \end{subfigure}
  \hfill
  \begin{subfigure}[b]{0.48\textwidth}
    \includegraphics[width=\textwidth]{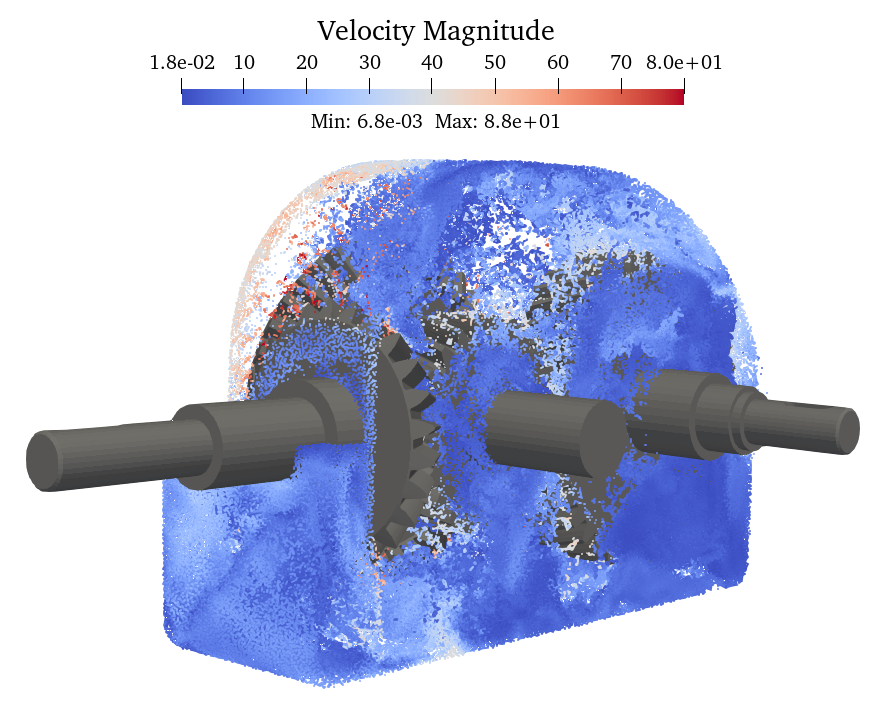}
    \caption{Oil-depth = -0.036 \textbackslash $\nu$ = 44 \textbackslash $\omega$ = 600}
    \label{fig:velocity-Oil-depth-0.036-44-600}
  \end{subfigure}
  
  \caption{The splash and velocity field of the lubricant when Shaft3 rotates 5 circles under different working conditions}
  \label{fig:Oil-splash-2-circles}
\end{figure}

\begin{figure}[htbp]
  \centering
  \begin{subfigure}[b]{0.48\textwidth}
    \includegraphics[width=\textwidth]{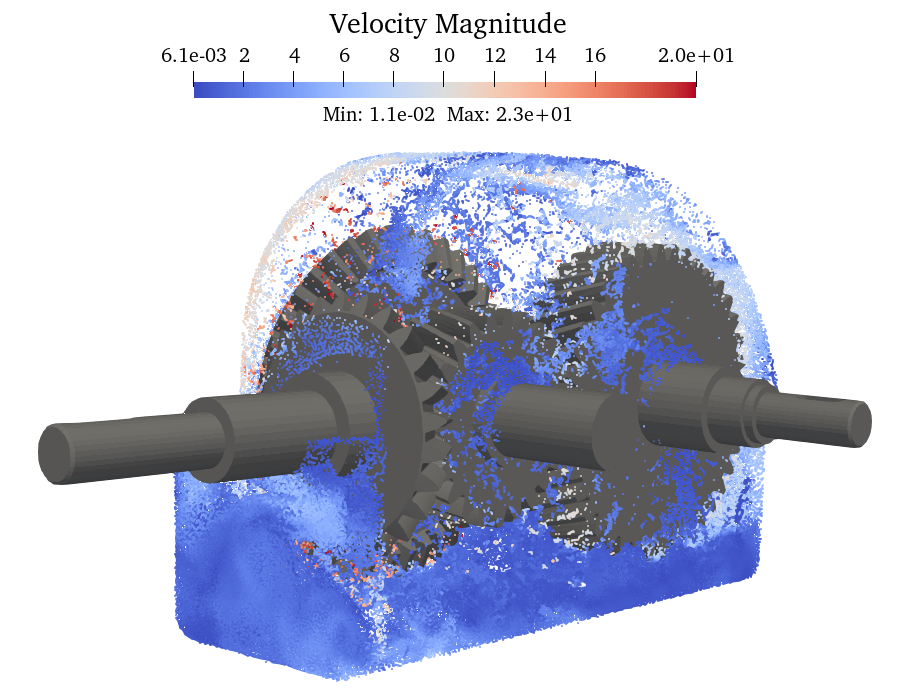}
    \caption{Oil-depth = -0.05 \textbackslash $\nu$ = 15 \textbackslash $\omega$ = 150 }
    \label{fig:velocity-Oil-depth-0.05-15-150N10}
  \end{subfigure}
  \hfill
  \begin{subfigure}[b]{0.48\textwidth}
    \includegraphics[width=\textwidth]{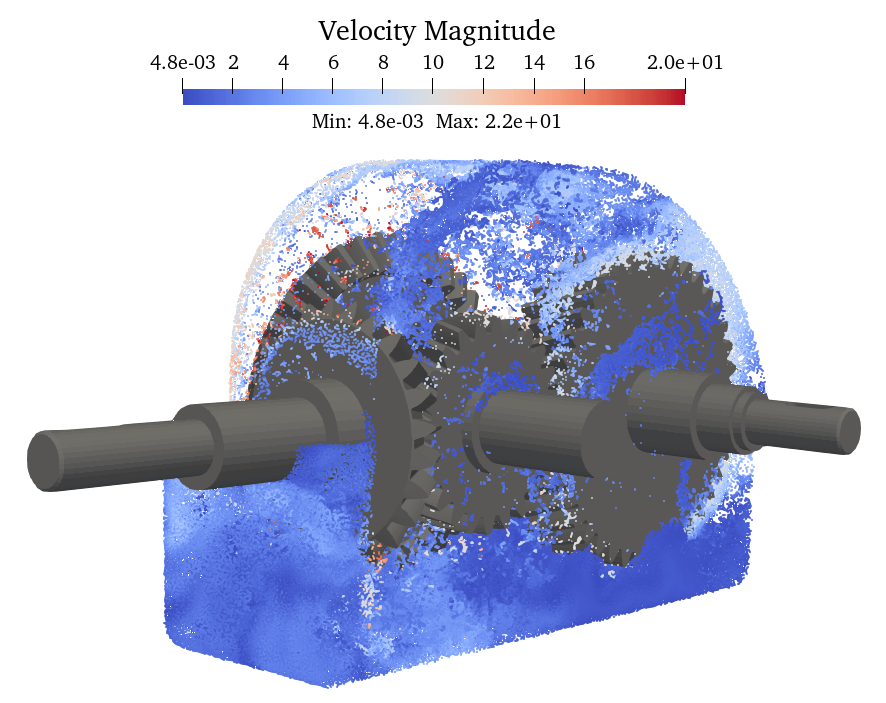}
    \caption{Oil-depth = -0.05 \textbackslash $\nu$ = 44 \textbackslash $\omega$ = 150 }
    \label{fig:velocity-Oil-depth-0.05-44-150N10}
  \end{subfigure}
  \vspace{0.3cm} 
  
  \begin{subfigure}[b]{0.48\textwidth}
    \includegraphics[width=\textwidth]{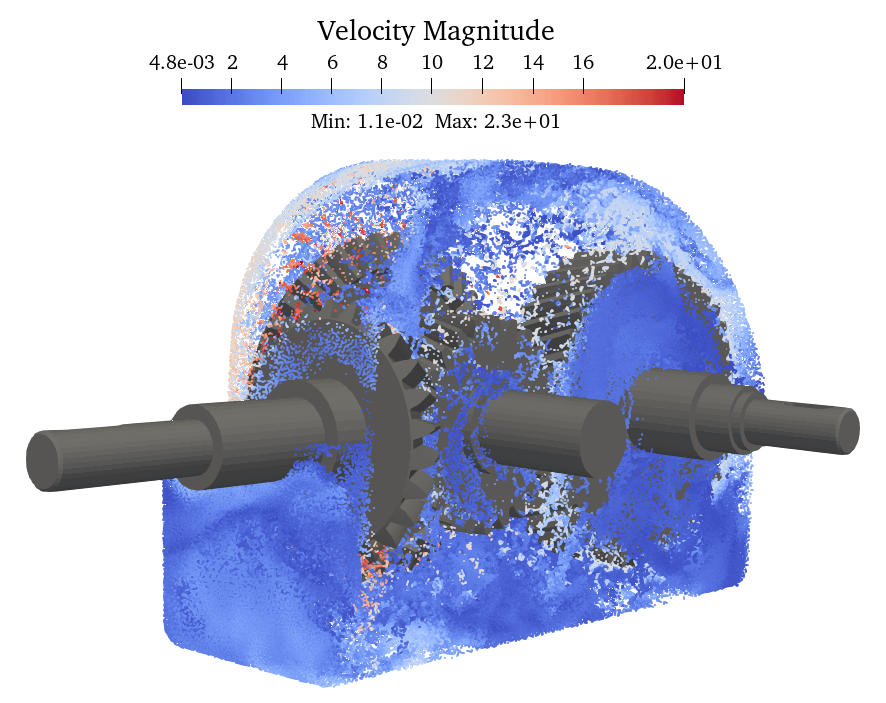}
    \caption{Oil-depth = -0.036 \textbackslash $\nu$ = 15 \textbackslash $\omega$ = 150 }
    \label{fig:velocity-Oil-depth-0.036-15-150N10}
  \end{subfigure}
  \hfill
  \begin{subfigure}[b]{0.48\textwidth}
    \includegraphics[width=\textwidth]{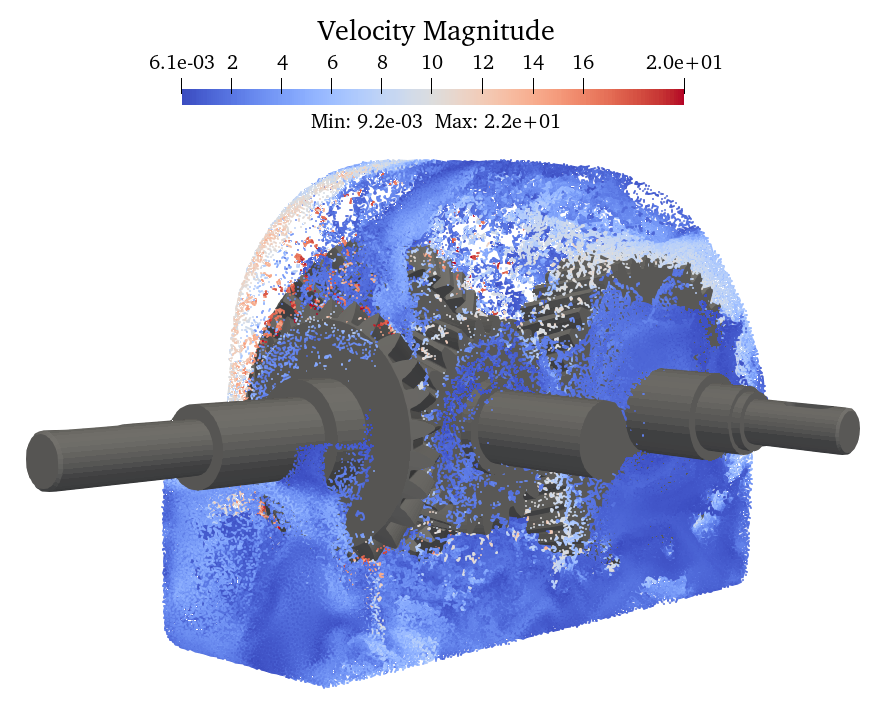}
    \caption{Oil-depth = -0.036 \textbackslash $\nu$ = 44 \textbackslash $\omega$ = 150 }
    \label{fig:velocity-Oil-depth-0.036-44-150N10}
  \end{subfigure}

  \begin{subfigure}[b]{0.48\textwidth}
    \includegraphics[width=\textwidth]{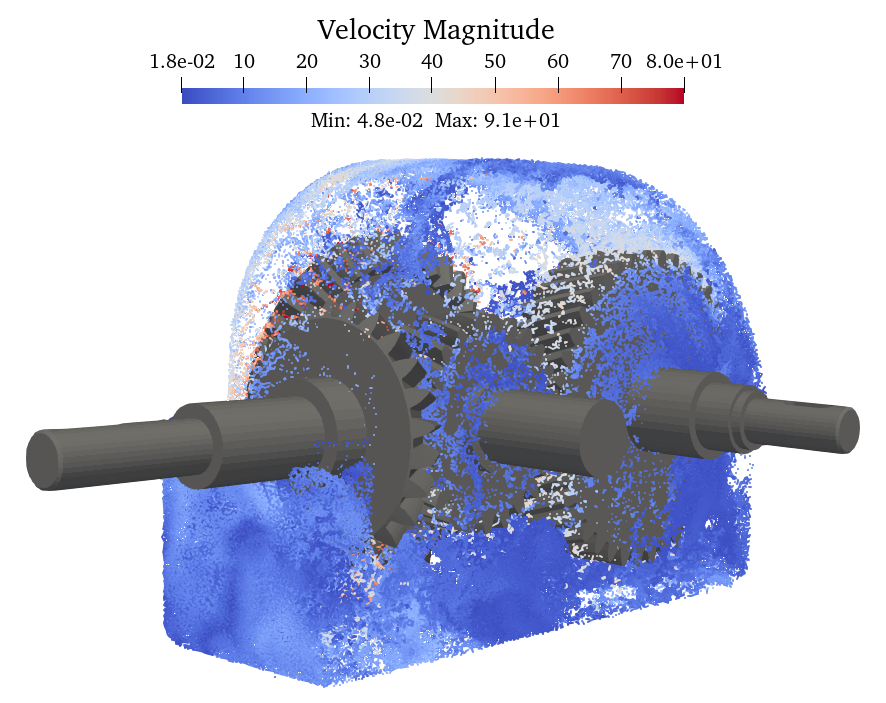}
    \caption{Oil-depth = -0.05 \textbackslash $\nu$ = 15 \textbackslash $\omega$ = 600}
    \label{fig:velocity-Oil-depth-0.05-15-600N10}
  \end{subfigure}
  \hfill
  \begin{subfigure}[b]{0.48\textwidth}
    \includegraphics[width=\textwidth]{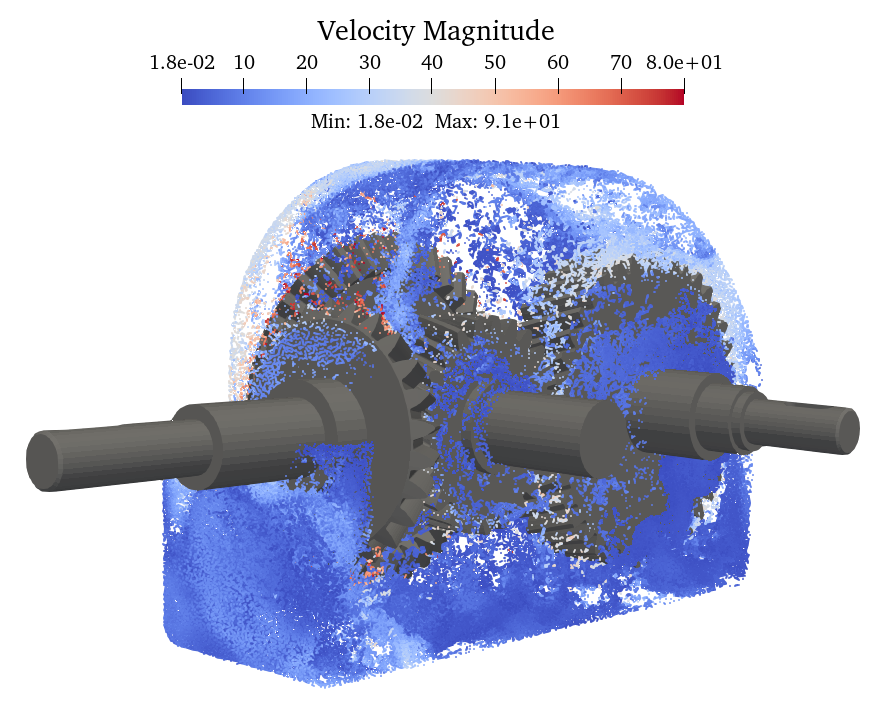}
    \caption{Oil-depth = -0.05 \textbackslash $\nu$ = 44 \textbackslash $\omega$ = 600}
    \label{fig:velocity-Oil-depth-0.05-44-600N10}
  \end{subfigure}
  \vspace{0.3cm} 
  
  \begin{subfigure}[b]{0.48\textwidth}
    \includegraphics[width=\textwidth]{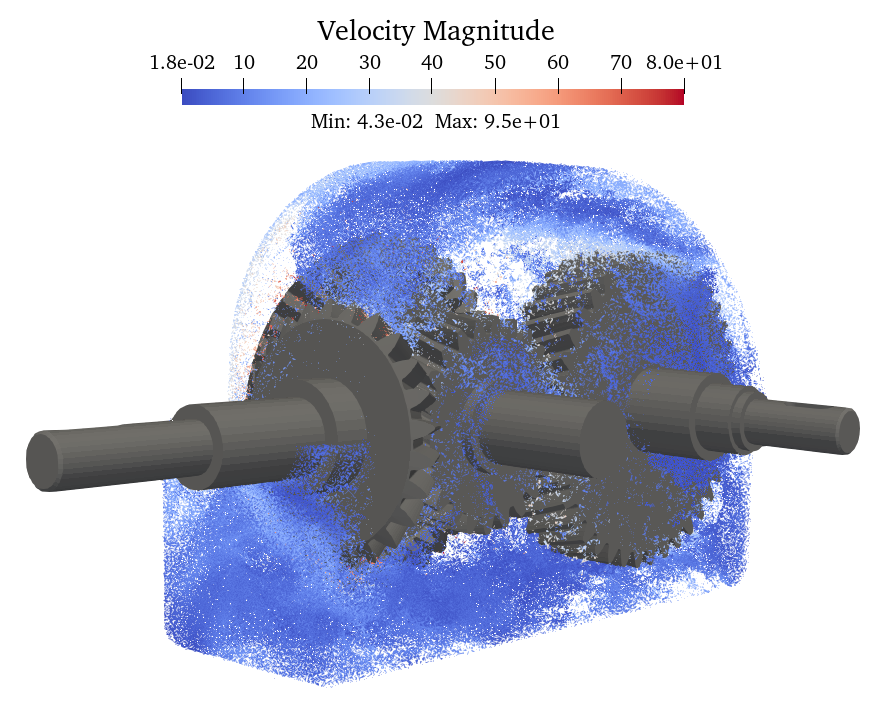}
    \caption{Oil-depth = -0.036 \textbackslash $\nu$ = 15 \textbackslash $\omega$ = 600}
    \label{fig:Oil-depth-0.036-15-600N10}
  \end{subfigure}
  \hfill
  \begin{subfigure}[b]{0.48\textwidth}
    \includegraphics[width=\textwidth]{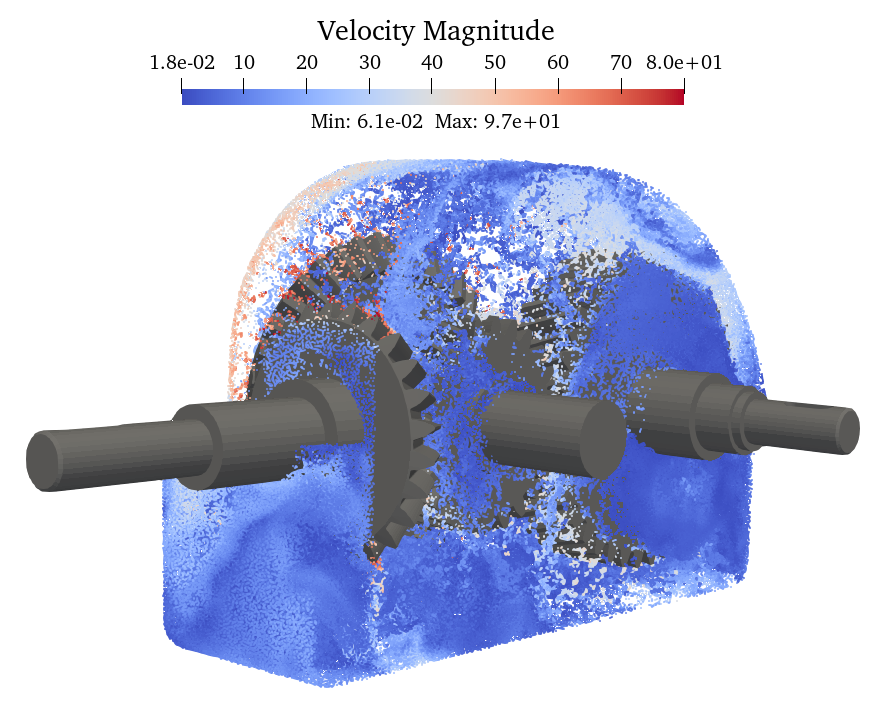}
    \caption{Oil-depth = -0.036 \textbackslash $\nu$ = 44 \textbackslash $\omega$ = 600}
    \label{fig:Oil-depth-0.036-44-600N10}
  \end{subfigure}
  
  \caption{The splash and velocity field of lubricant when Shaft3 rotates 10 circles under different working conditions.
}
  \label{fig:Oil-splash-velocity-10-circles}
\end{figure}

To further examine the transient behavior of the lubricant, Figures \ref{fig:Oil-splash-2-circles} 
and \ref{fig:Oil-splash-velocity-10-circles} visualize the splash morphology and velocity distribution of the 
lubricant at two distinct stages: after 5 and 10 revolutions of Shaft 3. 
To enhance the visibility of internal flow structures, only lubricant particles located 
beyond a slicing plane 0.01 m from the gearbox center along the positive $x$-axis are retained.
While the colormap range for velocity magnitude is kept consistent within groups of cases sharing 
the same shaft speed to facilitate visual comparison, the actual maximum velocity is annotated 
in each plot to reflect differences arising from specific operating conditions.

The splash patterns in both figures show relatively similar global structures at the two time points, 
suggesting that the lubricant dynamics reach a quasi-steady regime during the simulation. 
The lubricant consistently spreads throughout the gearbox and flows from the rear toward the front, 
eventually impinging on Bevel Gears 1 and 2. Higher local velocities are observed near their meshing area, 
indicating intense oil–gear interactions.
Additionally, due to the combined effects of the churning motion of Helical Gear 3 and the housing geometry, 
a noticeable amount of lubricant accumulates in the cavity beneath Shaft 1, forming a semi-stagnant region.
In contrast, Gears 3 and 4, rotating in opposite directions, actively stir and redistribute lubricant 
throughout the housing.

Among the observed cases, variations in lubricant viscosity do not substantially alter the splash morphology or velocity field. However, comparing cases \ref{fig:velocity-Oil-depth-0.036-15-600} 
and \ref{fig:velocity-Oil-depth-0.036-44-600} reveals that a higher viscosity slightly reduces the peak 
splash velocity. Interestingly, when comparing the high fill level cases at high rotational speed, 
the effect of viscosity becomes more pronounced: greater viscosity results in a larger amount of lubricant 
reaching higher velocities, as evidenced by the broader red regions in the plots.

While the effect of lubricant volume is less apparent during the early stages (e.g., Shaft 3 at 5 revolutions), 
the impact becomes more significant in the later stages (around 10 revolutions of Shaft 3), 
where higher oil fill levels produce more extensive coverage and encapsulation of the gear assembly. 
This suggests that additional lubricant gradually distributes throughout the housing, 
aided by sustained gear motion, and improves the overall lubrication coverage even under high-speed operation.

\begin{figure}[htbp]
  \centering
  
  \begin{subfigure}[b]{0.45\textwidth}
    \includegraphics[width=\textwidth]{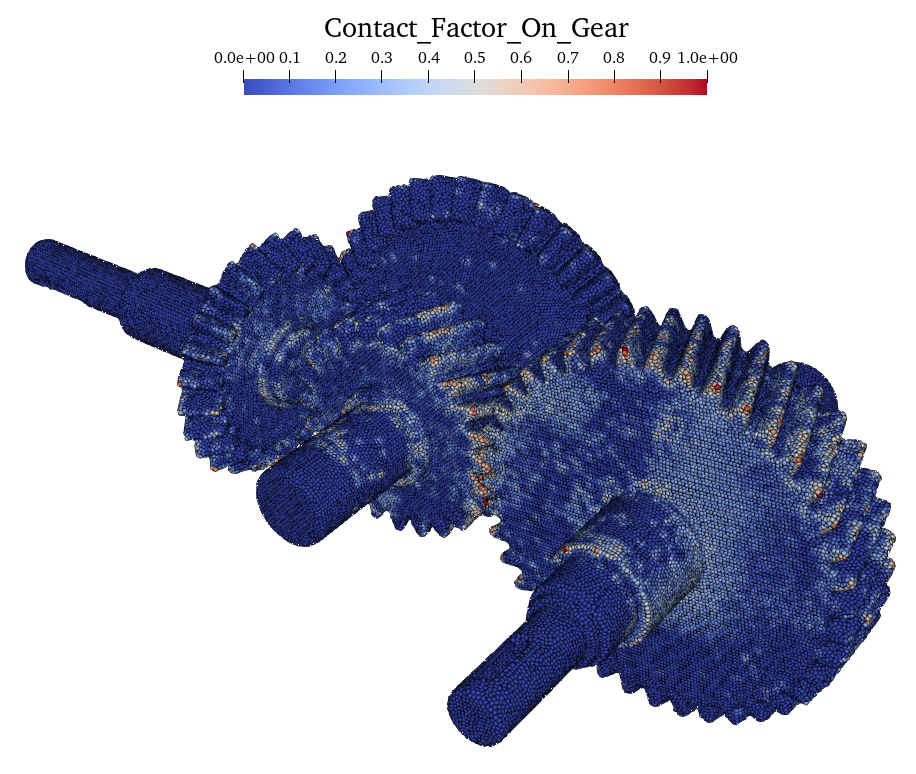}
    \caption{Oil-depth = -0.05 \textbackslash $\nu$ = 15 \textbackslash $\omega$ = 150 }
    \label{fig:CF-Oil-depth-0.05-15-150N2}
  \end{subfigure}
  \hfill
  \begin{subfigure}[b]{0.45\textwidth}
    \includegraphics[width=\textwidth]{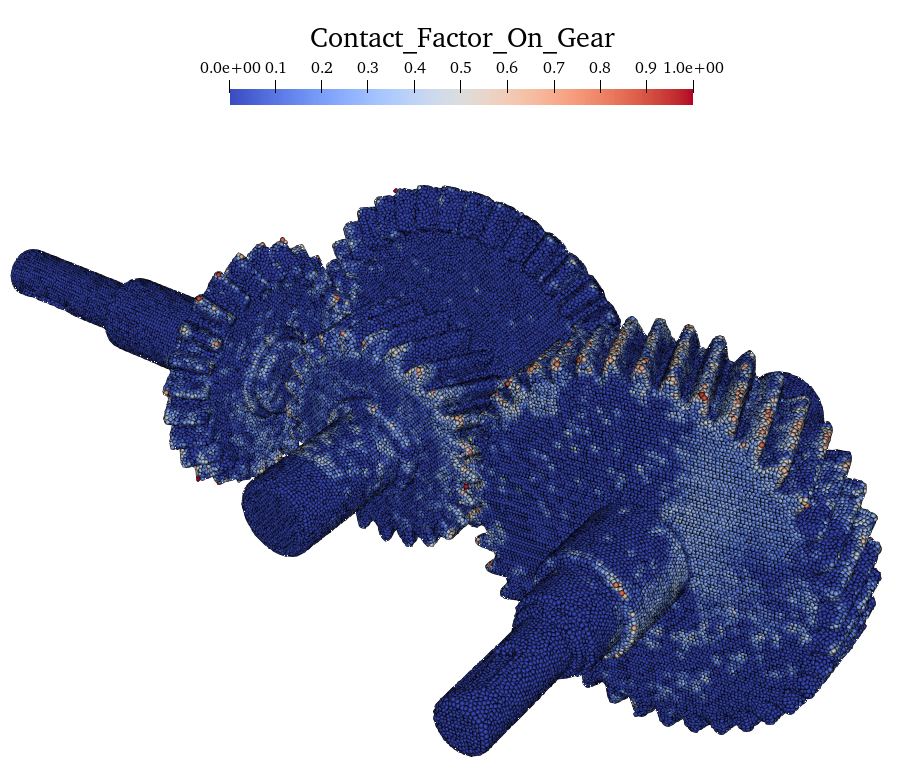}
    \caption{Oil-depth = -0.05 \textbackslash $\nu$ = 44 \textbackslash $\omega$ = 150 }
    \label{fig:OC-Fil-depth-0.05-44-150N2}
  \end{subfigure}
  \vspace{0.3cm} 
  
   \begin{subfigure}[b]{0.45\textwidth}
    \includegraphics[width=\textwidth]{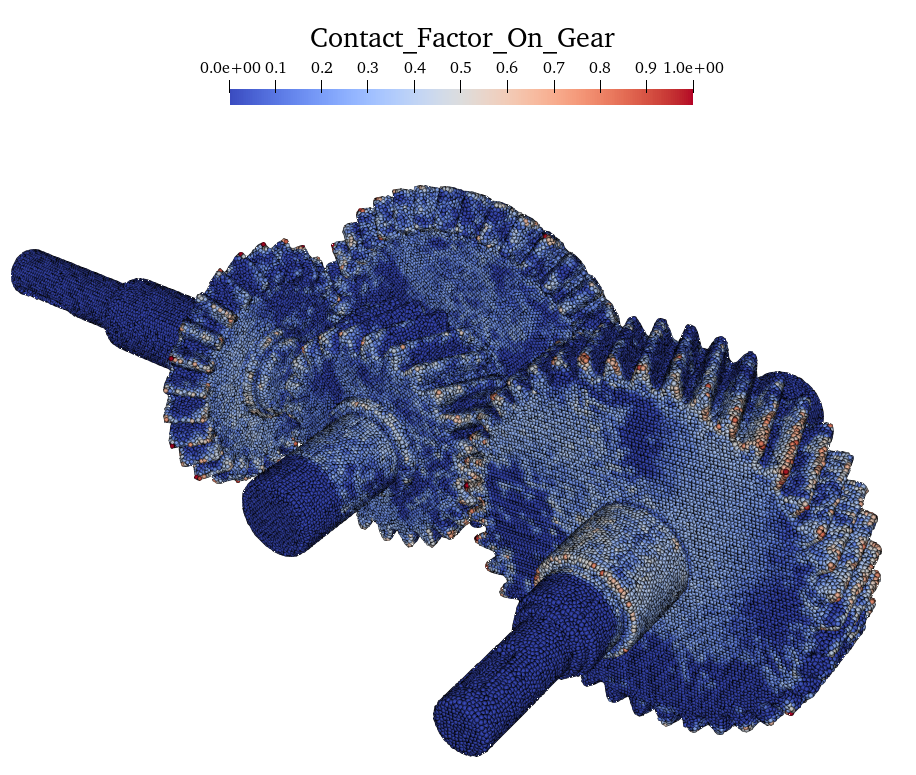}
    \caption{Oil-depth = -0.036 \textbackslash $\nu$ = 15 \textbackslash $\omega$ = 150 }
    \label{fig:CF-Oil-depth-0.05-15-150}
  \end{subfigure}
  \hfill
  \begin{subfigure}[b]{0.45\textwidth}
    \includegraphics[width=\textwidth]{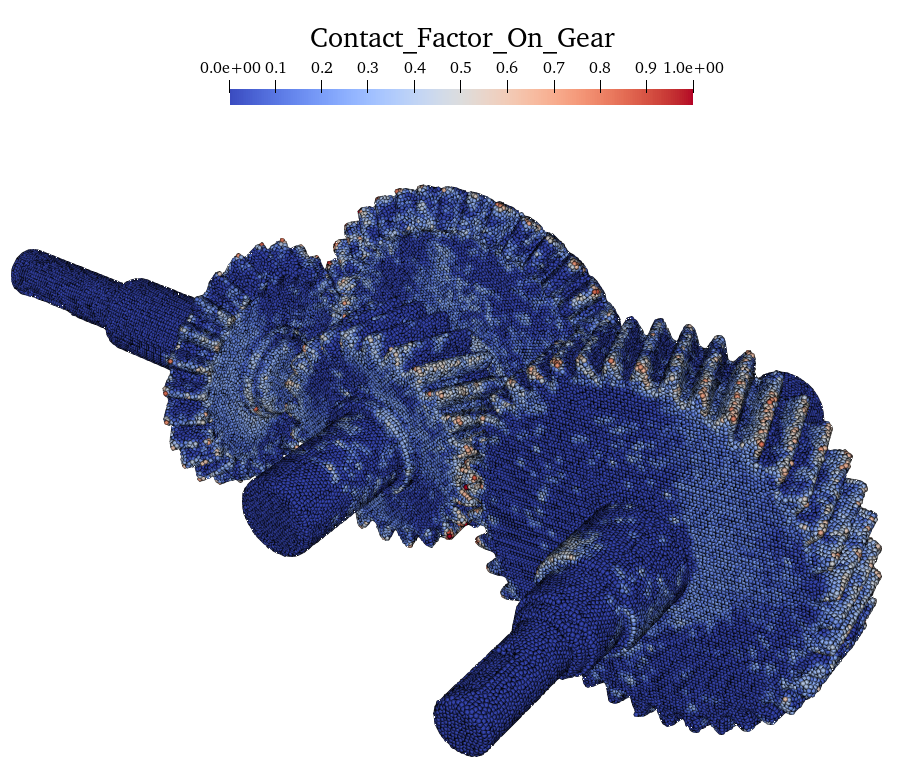}
    \caption{Oil-depth = -0.036 \textbackslash $\nu$ = 44 \textbackslash $\omega$ = 150 }
    \label{fig:CF-Oil-depth-0.05 - 44 -50}
  \end{subfigure}
  \vspace{0.3cm} 
  
 \begin{subfigure}[b]{0.45\textwidth}
    \includegraphics[width=\textwidth]{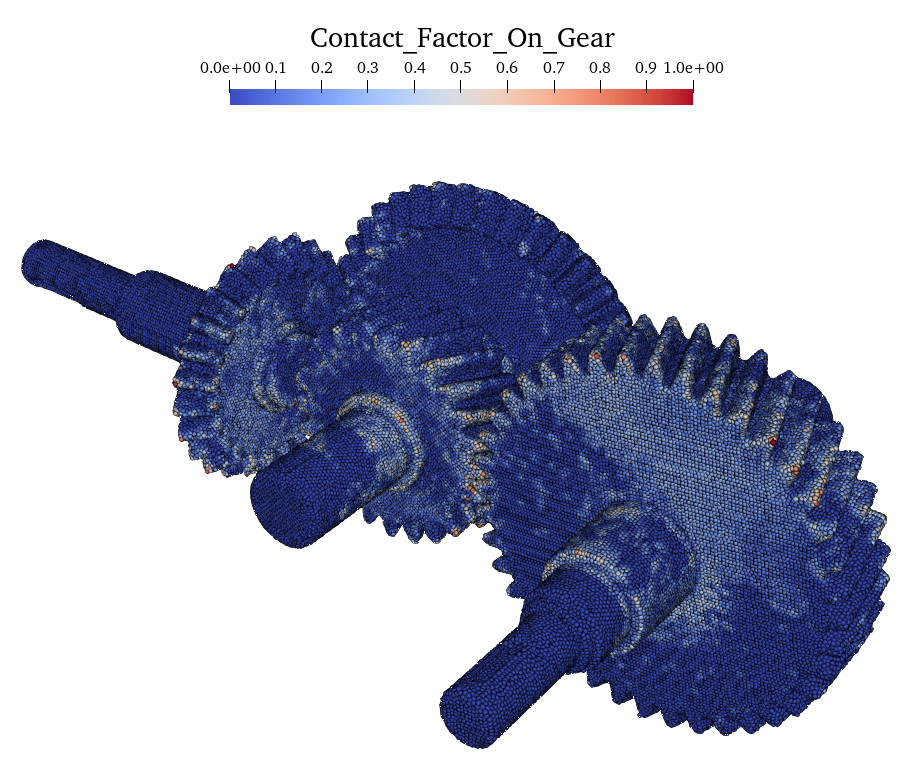}
    \caption{Oil-depth = -0.05 \textbackslash $\nu$ = 15 \textbackslash $\omega$ = 600 }
    \label{fig:CF-Oil-depth-0.05-15-600N2}
  \end{subfigure}
  \hfill
  \begin{subfigure}[b]{0.45\textwidth}
    \includegraphics[width=\textwidth]{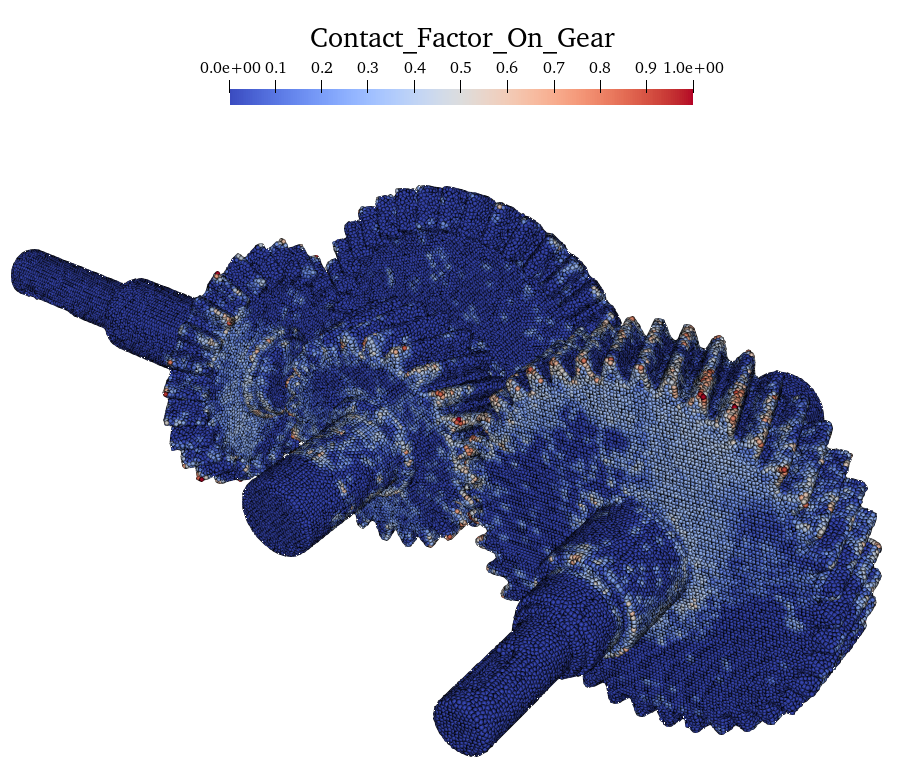}
    \caption{Oil-depth = -0.005 \textbackslash $\nu$ = 44 \textbackslash $\omega$ = 600 }
    \label{fig:CF-Oil-depth-0.005-44-600N2}
  \end{subfigure}
  
  \begin{subfigure}[b]{0.45\textwidth}
    \includegraphics[width=\textwidth]{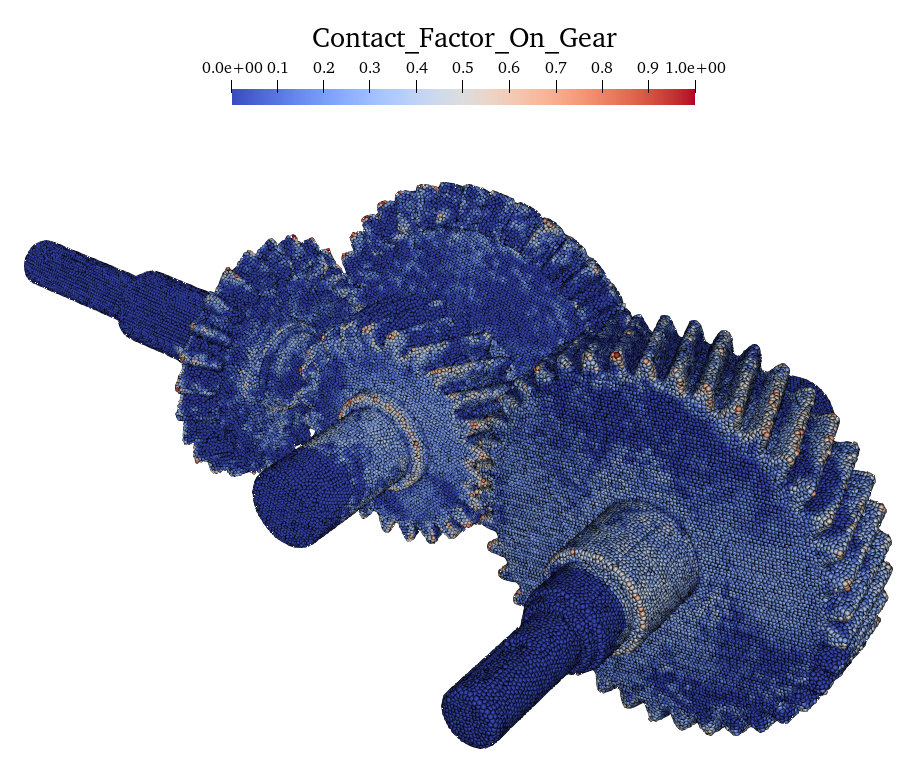}
    \caption{Oil-depth = -0.036 \textbackslash $\nu$ = 15 $\omega$ = 600 }
    \label{fig:CF-Oil-depth-0.036-15-150}
  \end{subfigure}
  \hfill
  \begin{subfigure}[b]{0.45\textwidth}
    \includegraphics[width=\textwidth]{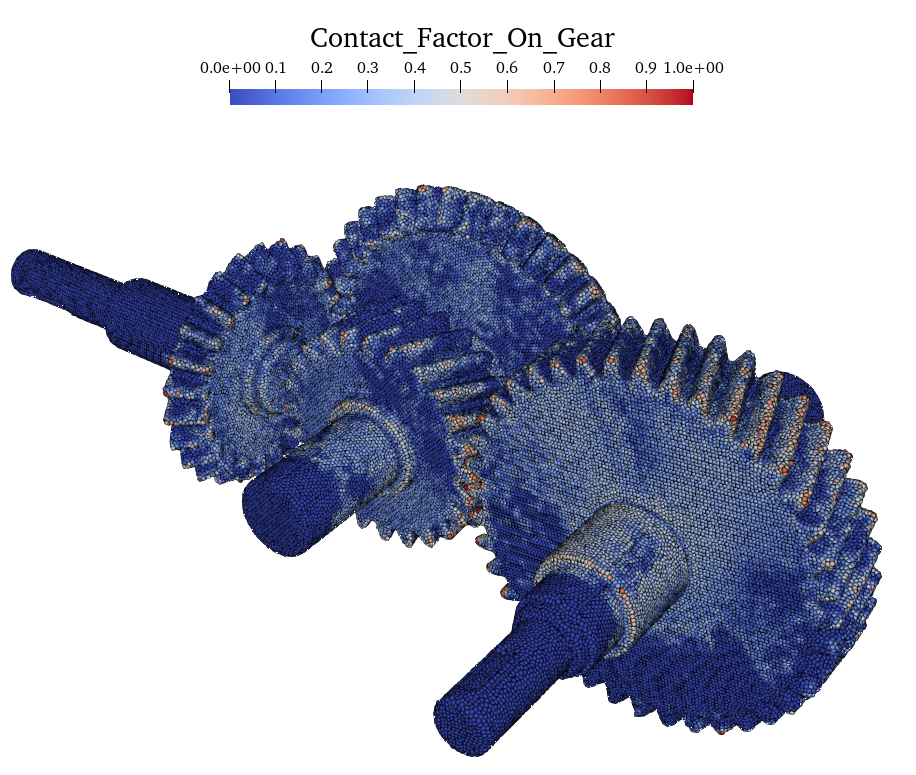}
    \caption{Oil-depth = -0.036 \textbackslash $\nu$ = 44 \textbackslash $\omega$ = 600 }
    \label{fig:CF-Oil-depth-0.036-44-150N2}
  \end{subfigure}
  
  \caption{Contact factor distribution under different operating conditions when Shaft3 rotates 2 circles.
}
  \label{fig:Contact-factor-2-circles}
\end{figure}

\begin{figure}[htbp]
  \centering
  \begin{subfigure}[b]{0.45\textwidth}
    \includegraphics[width=\textwidth]{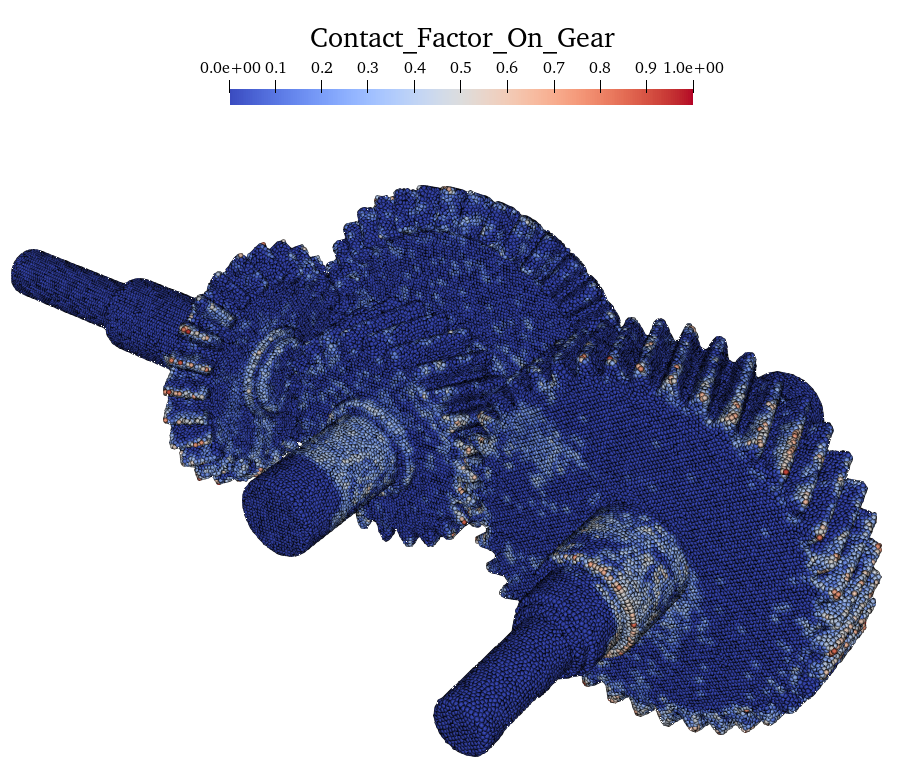}
    \caption{Oil-depth = -0.05 \textbackslash $\nu$ = 15 \textbackslash $\omega$ = 150 }
    \label{fig:CF-Oil-depth-0.05-15-150N10}
  \end{subfigure}
  \hfill
  \begin{subfigure}[b]{0.45\textwidth}
    \includegraphics[width=\textwidth]{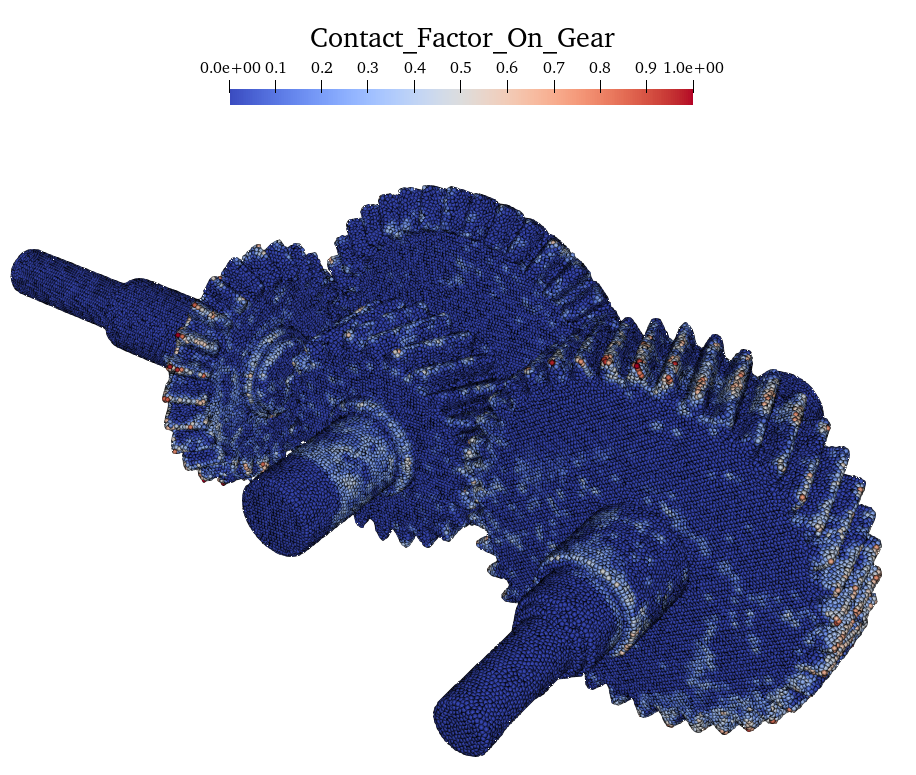}
    \caption{Oil-depth = -0.05 \textbackslash $\nu$ = 44 \textbackslash $\omega$ = 150 }
    \label{fig:CF-Oil-depth-0.05-44-150N10}
  \end{subfigure}
  \vspace{0.3cm} 

  \begin{subfigure}[b]{0.45\textwidth}
    \includegraphics[width=\textwidth]{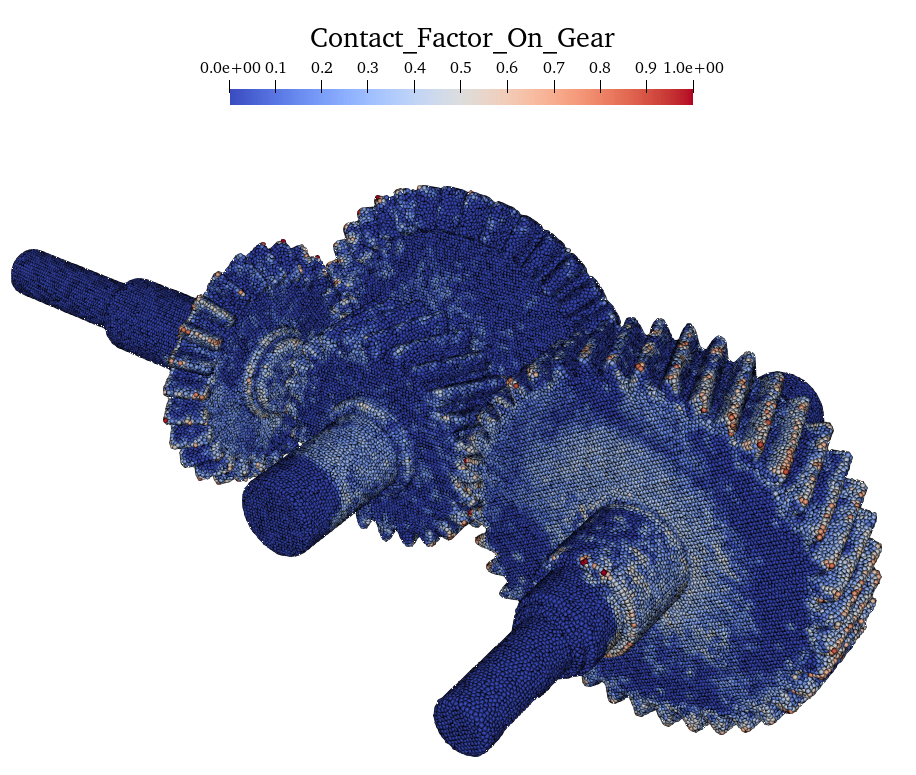}
    \caption{Oil-depth = -0.036 \textbackslash $\nu$ = 15 \textbackslash $\omega$ = 150 }
    \label{fig:CF-Oil-depth-0.05-15 /150}
  \end{subfigure}
  \hfill
  \begin{subfigure}[b]{0.45\textwidth}
    \includegraphics[width=\textwidth]{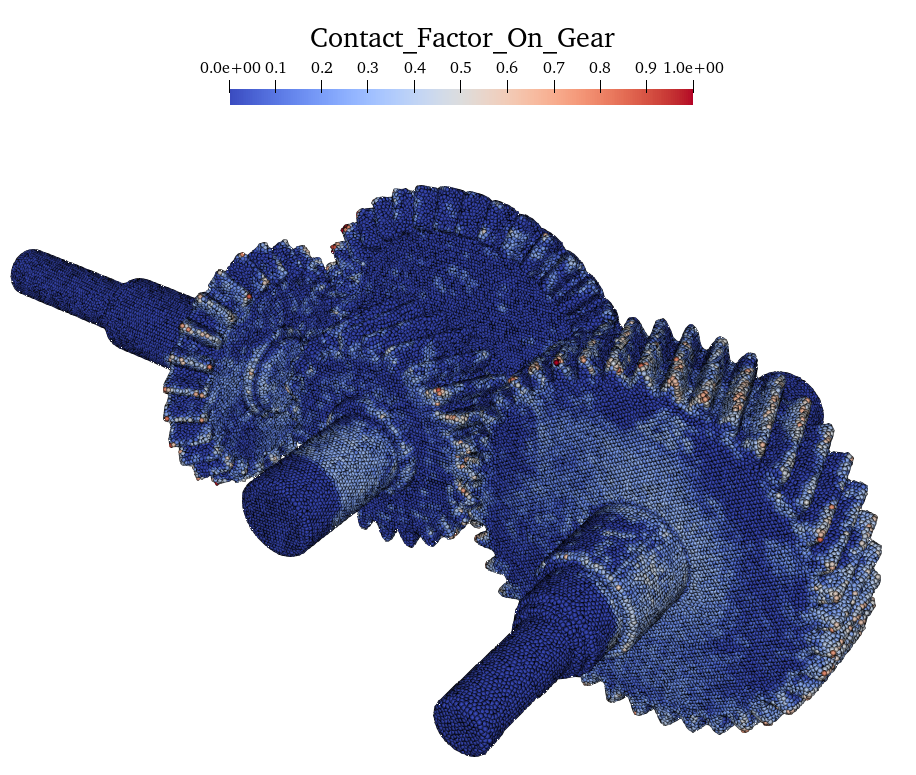}
    \caption{Oil-depth = -0.036 \textbackslash $\nu$ = 44 \textbackslash $\omega$ = 150 }
    \label{fig:CF-Oil-depth-0.05-44-150N10}
  \end{subfigure}
  \vspace{0.3cm} 
  
  \begin{subfigure}[b]{0.45\textwidth}
    \includegraphics[width=\textwidth]{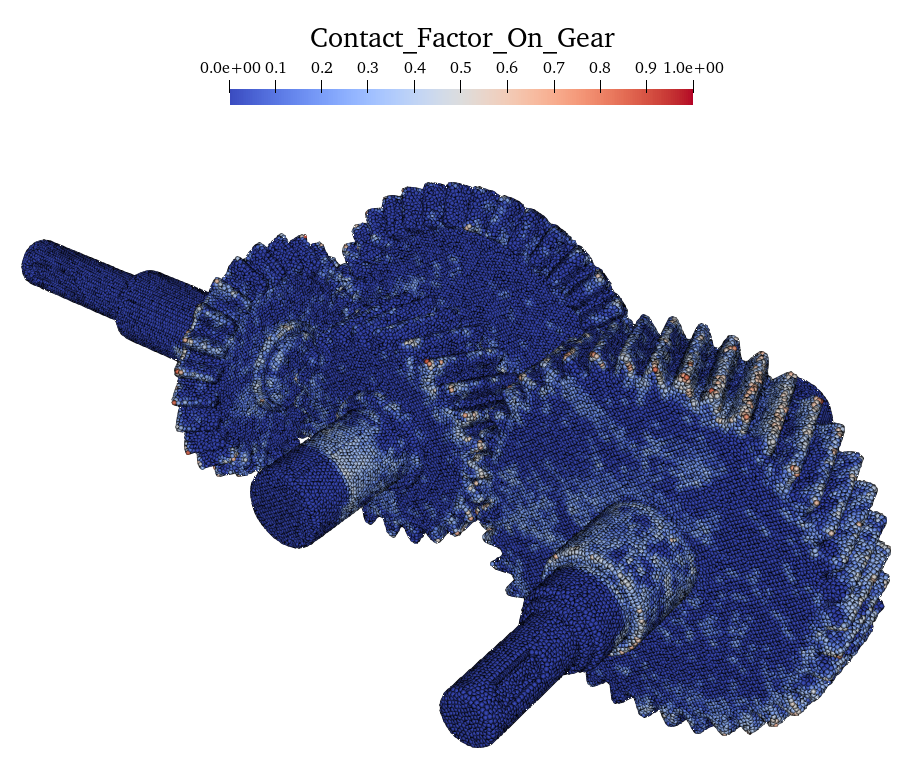}
    \caption{Oil-depth = -0.05 \textbackslash $\nu$ = 15 \textbackslash $\omega$ = 600 }
    \label{fig:CF-Oil-depth-0.036-15-150N10}
  \end{subfigure}
  \hfill
  \begin{subfigure}[b]{0.48\textwidth}
    \includegraphics[width=\textwidth]{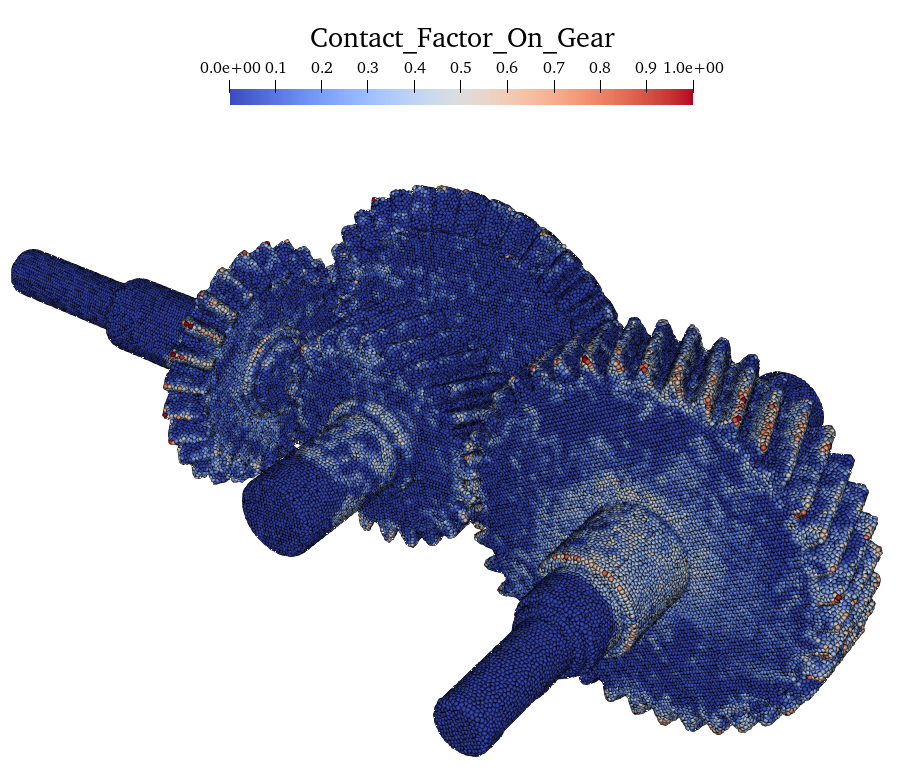}
    \caption{Oil-depth = -0.05 \textbackslash $\nu$ = 44 \textbackslash $\omega$ = 600 }
    \label{fig:CF-Oil-depth-0.036-44-150N10}
  \end{subfigure}

  \begin{subfigure}[b]{0.45\textwidth}
    \includegraphics[width=\textwidth]{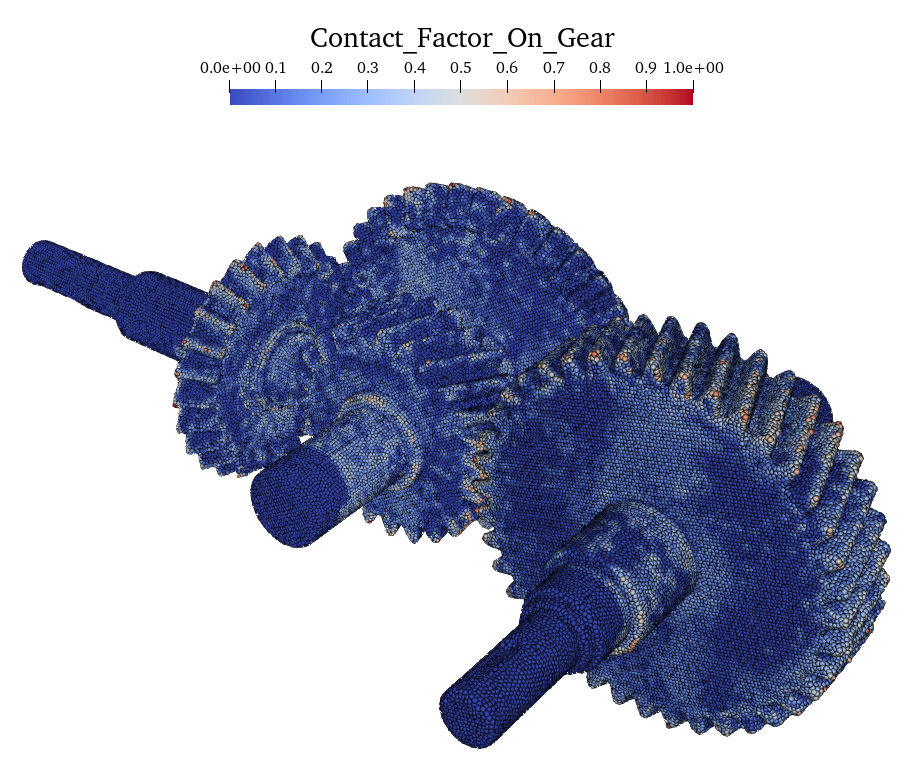}
    \caption{Oil-depth = -0.036 \textbackslash $\nu$ = 15 \textbackslash $\omega$ = 600 }
    \label{fig:CF-Oil-depth-0.036-15= 150N10}
  \end{subfigure}
  \hfill
  \begin{subfigure}[b]{0.45\textwidth}
    \includegraphics[width=\textwidth]{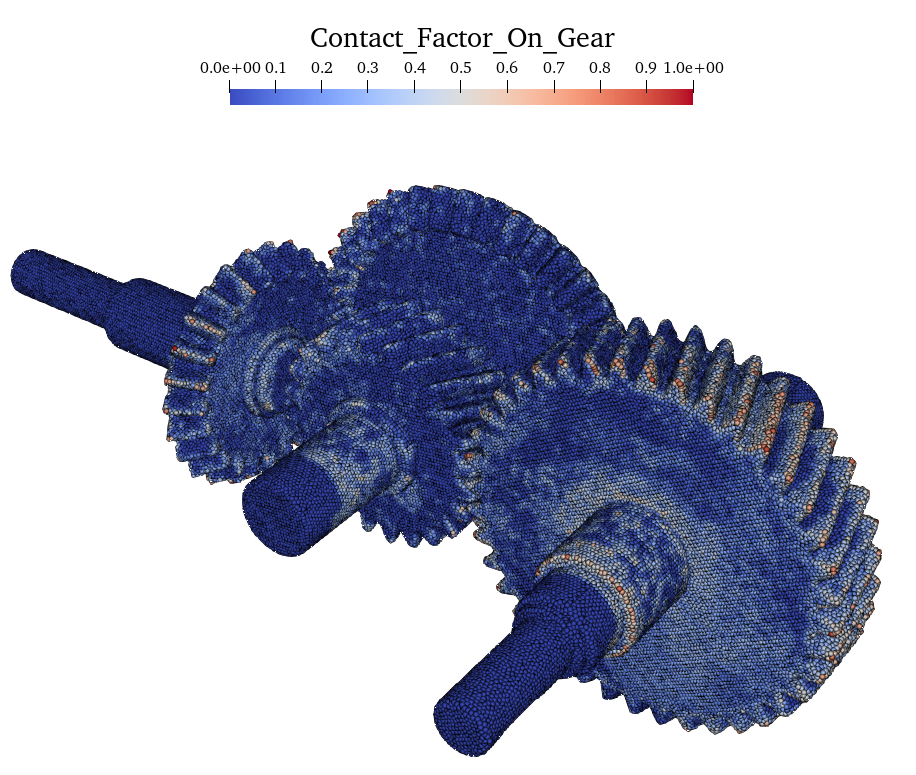}
    \caption{Oil-depth = -0.036 \textbackslash $\nu$ = 44 \textbackslash $\omega$ = 600 }
    \label{fig:CF-Oil-depth-0.036-44-150N10}
  \end{subfigure}
  
  \caption{Contact factor distribution under different operating conditions when Shaft3 rotates 10 circles.}
  \label{fig:Contact-factor-10-circless}
\end{figure}

Figures \ref{fig:Contact-factor-2-circles} and \ref{fig:Contact-factor-10-circless} show the distribution 
of the contact factor, a normalized indicator of the contact degree between lubricating oil and the gear 
(see Section \ref{subsec:contact-factor} for the definition). 
These figures further analyze the oil–gear interaction at early and later simulation stages, 
corresponding to 2 and 10 revolutions of Shaft 3, respectively.
Initially, contact is more concentrated on the gear flanks and shaft surfaces, 
with less engagement across the gear teeth. As the simulation progresses, contact becomes more 
evenly distributed, particularly across the entire gear tooth profiles, 
suggesting improved oil circulation and coverage over time.

Notably, at the early simulation stages, the contact factors observed along the 
gear circumference are predominantly concentrated in the meshing regions, 
likely reflecting the intended splash targets shaped by the gearbox design. 
This observation aligns with the lubricant splash patterns previously shown 
in Figures \ref{fig:Oil-splash-2-circles} and \ref{fig:Oil-splash-velocity-10-circles}, 
further supporting the consistency between oil trajectory and contact behavior.
In later stages, the contact factor distribution becomes more uniform, 
thereby reducing the risk of dry contact and associated wear. 
Across all scenarios, lubricant viscosity shows minimal influence on the contact factor, 
whereas oil volume and shaft speed exert more pronounced effects. 
Higher oil levels improve surface coverage, while increased rotational speeds 
facilitate a more even and dynamic redistribution of lubricant throughout the system.

\begin{figure}[htbp]
  \centering
  \begin{subfigure}[b]{0.48\textwidth}
    \includegraphics[width=\textwidth]{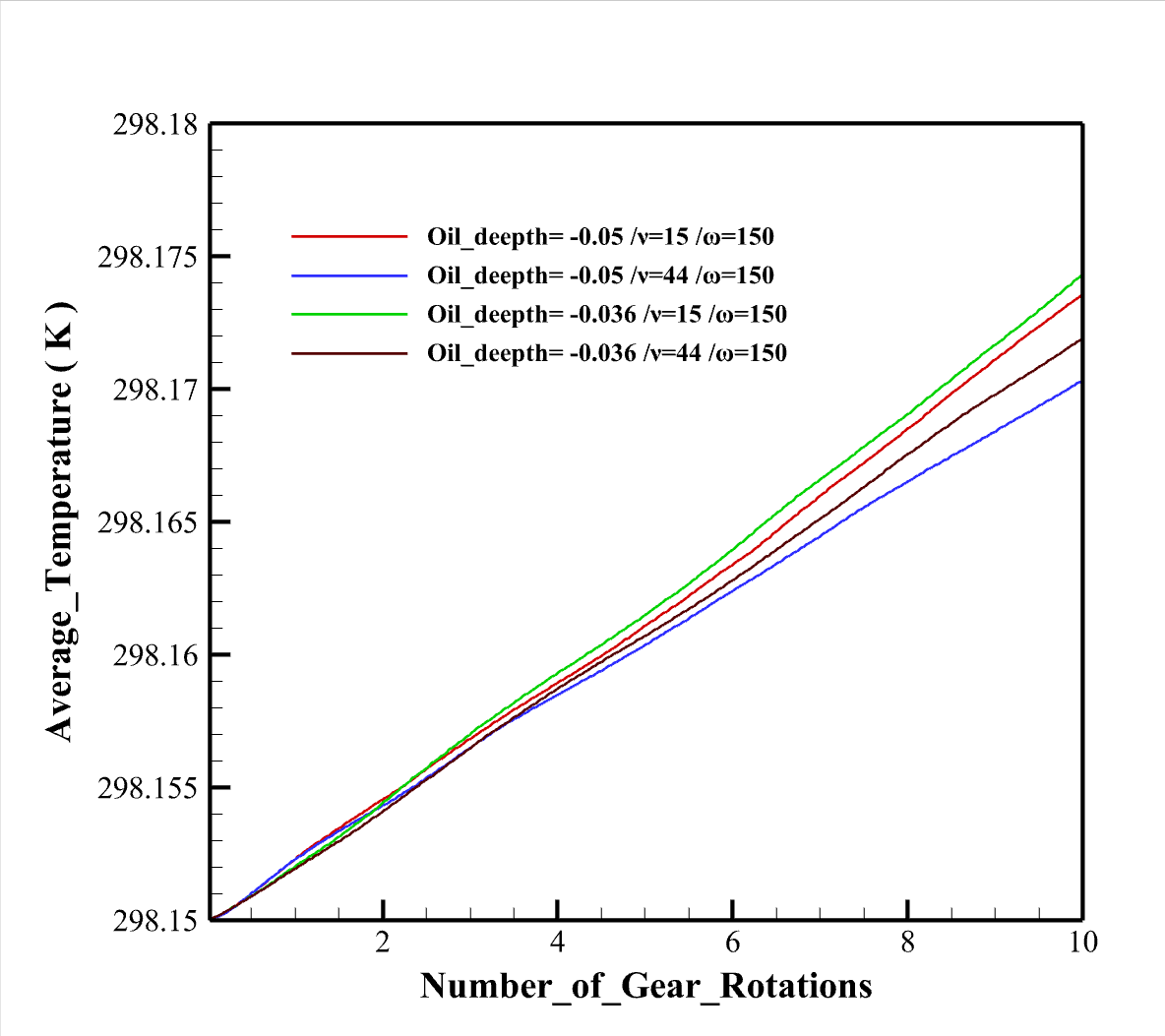}
    \caption{ }
    \label{fig:CF-Oil-depth-0.036-15= 150N10}
  \end{subfigure}
  \hfill
  \begin{subfigure}[b]{0.48\textwidth}
    \includegraphics[width=\textwidth]{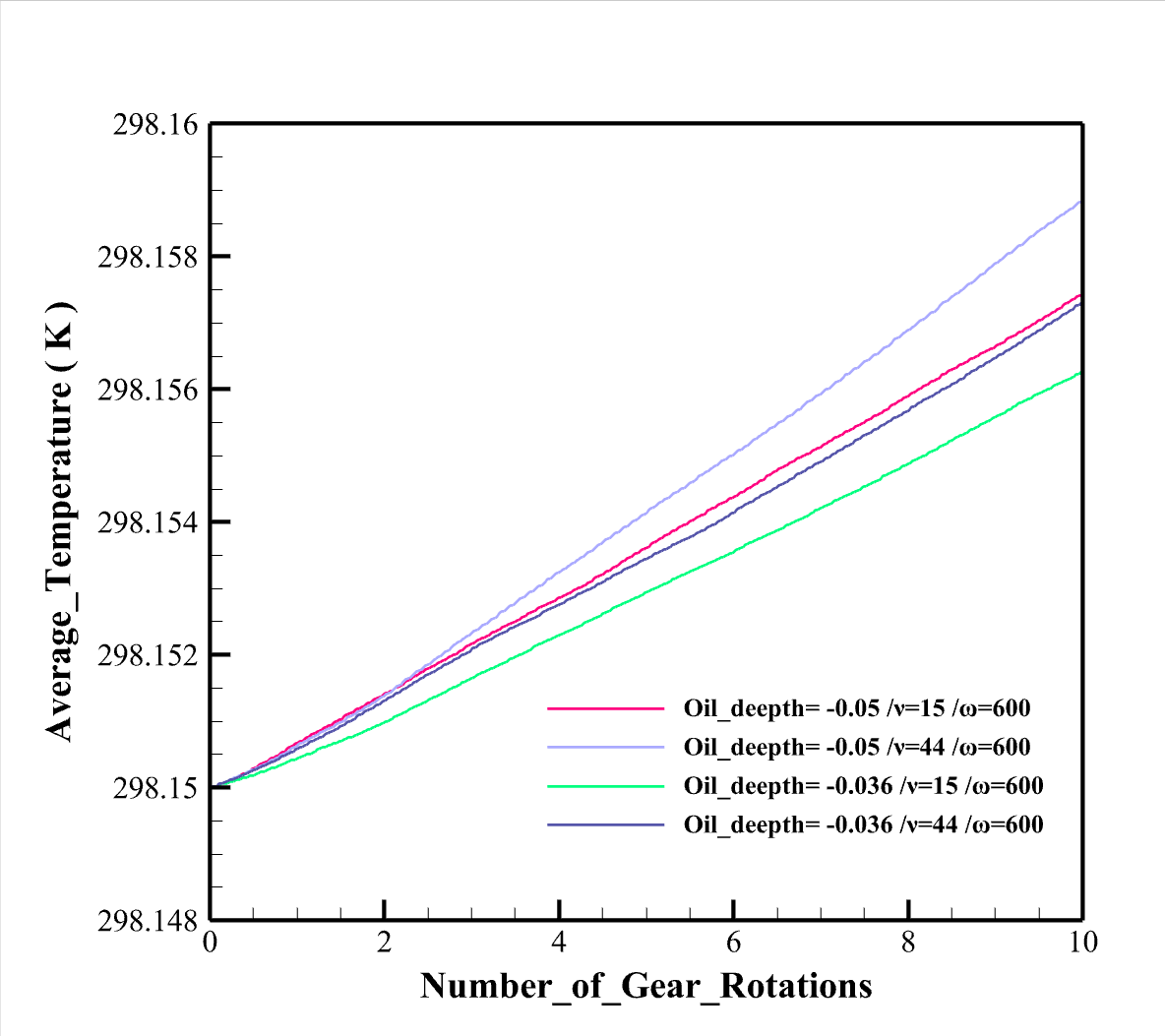}
    \caption{ }
    \label{fig:CF-Oil-depth-0.036-44-150N10}
  \end{subfigure}
  
  \caption{Figure 10: Evolution of the average lubricant temperature with respect to the 
  number of Shaft 3 rotations under two angular velocities: (a) $\omega$ = 150 rad/s; 
  (b) $\omega$ = 600 rad/s. Each subplot includes four cases with different oil viscosities and oil depths.}
  \label{fig:average-temperature}
\end{figure}

In this study, thermal simulations were conducted with the initial lubricant and gear temperatures 
set to 298.15 $\mathrm{K}$ and 318.15 $\mathrm{K}$, respectively. The gear temperature was assumed to be spatially 
uniform and held constant throughout the simulation. Heat generated by gear meshing or viscous 
dissipation was not considered; thus, the only heat transfer mechanism was thermal conduction 
and convection between the gear surfaces and the surrounding lubricant. The evolution of average 
lubricant temperature, as shown in Figure \ref{fig:average-temperature}, 
reflects solely the net heat absorbed from the gears under different operating conditions.

The results show that under lower angular velocity (150 $\mathrm{rad/s}$), the lubricant 
exhibits a slightly higher average temperature rise compared to the higher-speed case (600 $\mathrm{rad/s}$), 
despite identical initial conditions and lubricant volume. Since shear-induced heating is excluded 
from the model, this phenomenon must be attributed to differences in how efficiently the lubricant 
receives heat from the gear surfaces. A plausible explanation lies in the contact dynamics 
between oil and gear surfaces. At higher speeds, although the lubricant is more widely dispersed, 
as seen in the broader contact factor distributions, the increased velocity may reduce the effective 
contact time with hot gear surfaces, thus limiting the heat absorbed per unit volume of lubricant. 
In contrast, at lower speeds, the lubricant remains in longer and more stable contact with gear surfaces, 
allowing for more efficient thermal exchange and thus a slightly greater accumulation of heat.

The effects of oil volume (represented by oil depth) and viscosity are much weaker and, importantly,
exhibit opposite tendencies at low and high speeds as shown in Figure~\ref{fig:average-temperature}. 
At 150~$\mathrm{rad/s}$, deeper immersion (oil\_depth = $-0.036$~m) leads to a
slightly larger temperature rise than the shallower oil bath, 
whereas at 600~$\mathrm{rad/s}$ the trend reverses. 
This behavior indicates a shift between two regimes: 
at low speed the flow is relatively
weak and the dominant effect of increased oil depth is to expose more fluid to
the hot gear surfaces, so that a larger fraction of the lubricant repeatedly
recirculates through the heated region. At high speed the flow becomes much
more vigorous and the oil bath is more uniformly mixed, so that the additional
oil volume mainly acts as extra thermal capacity, 
resulting in a reduced average temperature rise.

The effect of viscosity shows a similar reversal. At 150~$\mathrm{rad/s}$, a
higher viscosity slightly suppresses convective mixing and limits the net heat
uptake by the bulk lubricant, leading to a smaller temperature increase.
At 600~$\mathrm{rad/s}$, 
although viscous dissipation is not modeled in this simulation, 
the enhanced viscosity may lead to more stable oil-gear contact and alter the extent of splashing, 
thereby enabling more sustained thermal exchange.
Thus, the more viscous oil maintains a more stable
oil--gear contact and thicker entrained films, which enhances the effective
thermal coupling between the gears and the fluid and yields a somewhat larger
temperature rise. 
This suggests that under high-speed conditions, 
moderate viscosity may facilitate more effective heat absorption from the gear.
It is worth noting that at high rotational speeds, the omission of gear-contact frictional heating 
and viscous dissipation causes the predicted average lubricant temperatures to represent lower-bound estimates.

In all cases, the depth- and viscosity-induced differences
remain within a few tens of percent and are much smaller than the strong
rotational speed dependence of the overall heating rate.
Taken together, these observations highlight the subtle but important interplay between 
fluid properties, gear motion, and thermal response, even in a simplified heat transfer model.

\subsection{Computational Performance Comparison on GPU and CPU Platforms}
\label{subsec:GPU-CPU}

To assess the computational viability of the proposed simulation framework 
in industrial-scale applications, two computationally intensive cases 
(Case 9 and Case 10 in Table \ref{table:operating-conditions}) are selected for 
benchmarking on CPU and GPU platforms. The hardware specifications of the tested 
systems are summarized in Table \ref{table:hardware-comparison}, which compares 
an RTX 2080 Super GPU and an Intel Core i9-11900 CPU. 
These platforms represent typical high-end consumer-grade devices that 
are accessible in industrial and research settings.

\begin{table}[htbp]
\centering
\caption{Key Specification Comparison: RTX 2080 Super vs. Intel Core i9-11900}
\label{table:hardware-comparison}
\begin{tabular}{lll}
\toprule
Specification & RTX 2080 Super & Core i9-11900 \\
\midrule
Architecture & Turing (TU104) & Rocket Lake \\
Cores / Threads & 3072 CUDA cores & 8 cores / 16 threads \\
Base Clock & 1650 MHz & 2.5 GHz \\
Boost Clock & 1815 MHz & Up to 5.2 GHz \\
Memory Type / Capacity & GDDR6 / 8 GB & System memory: 32 GB \\
\bottomrule
\end{tabular}
\end{table}

\begin{figure}[htbp]
  \centering
  \begin{subfigure}[b]{0.48\textwidth}
    \includegraphics[width=\textwidth]{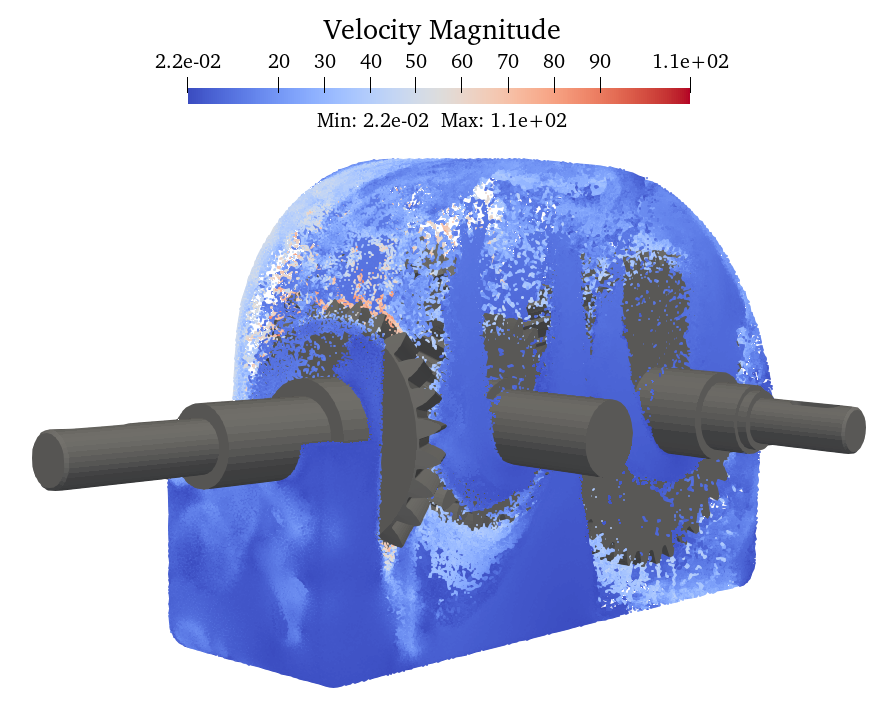}
    \caption{Oil-depth = -0.01 \textbackslash $\nu$ = 44 \textbackslash $\omega$ = 600 }
    \label{fig:5.7million-case}
  \end{subfigure}
  \hfill
  \begin{subfigure}[b]{0.48\textwidth}
    \includegraphics[width=\textwidth]{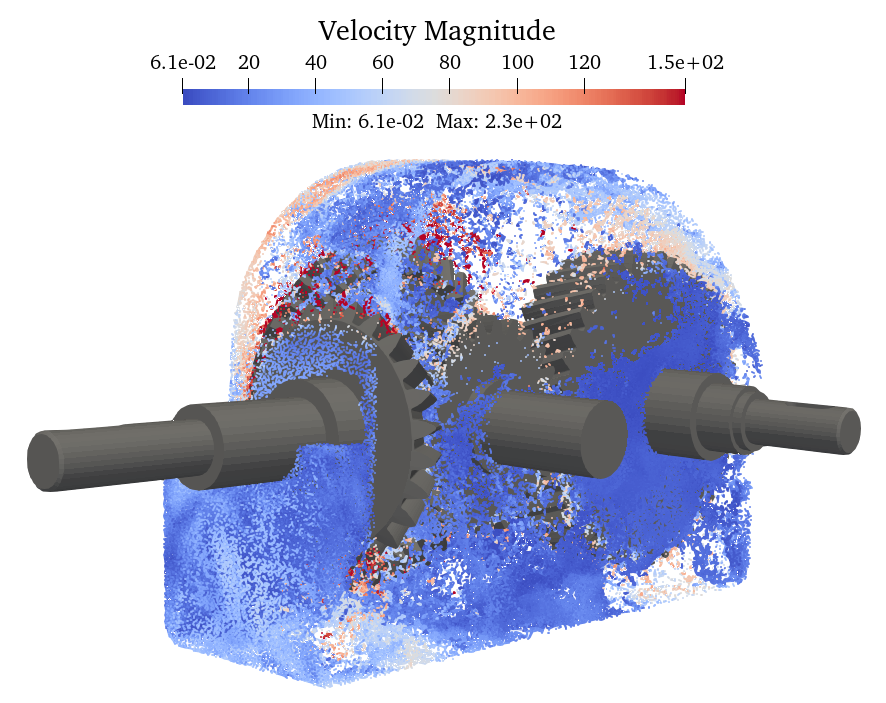}
    \caption{Oil-depth = -0.05 \textbackslash $\nu$ = 44 \textbackslash $\omega$ = 1500 }
    \label{fig:1500rad-case}
  \end{subfigure}
  \vspace{0.3cm} 
 \caption{velocity field of lubricant splashing at two test conditions.
 The higher resolution in case (a) enables more detailed visualization of the splash pattern.}
  \label{fig:velocity-comparison-cpu-gpu}
\end{figure}

The two benchmark cases represent distinct computational extremes. 
Case 9 involves a high shaft speed of 1500 $\mathrm{rad/s}$, resulting in a tip speed 
of 100.5 $\mathrm{m/s}$ for Gear 4. Due to intense oil–gear interaction, 
the simulated lubricant splash velocity reaches over 200 $\mathrm{m/s}$ (As shown in Fig. \ref{fig:1500rad-case}).
This necessitates a stringent time step of $2.0 \times 10^{-9}$ to ensure numerical 
stability and accurate resolution of rapid fluid transients.
Case 10, on the other hand, features a moderately high shaft speed of 600 $\mathrm{rad/s}$, 
coupled with a significantly elevated lubricant fill and a finer particle resolution. 
This configuration leads to the involvement of over 5.7 million particles in the 
simulation, significantly increasing the computational load per time step.
Although the shaft speed in this case allows a slightly longer time step of $4.0 \times 10^{-8}$s, 
it still falls within the regime of high temporal resolution.
The simulation is implemented using a SYCL-based heterogeneous framework with 
Unified Shared Memory (USM), which allows host and device to access shared memory spaces, 
simplifying data management and reducing transfer overhead. 
Nevertheless, the large particle count imposes substantial memory pressure and processing 
demands on the GPU. Even under USM, physical memory constraints and parallel interaction 
calculations, especially for dense fluid–structure interactions, remain computationally intensive. 
This makes Case 10 a strong benchmark for evaluating the performance of large-scale 
industrial SPH simulations.
Taken together, these two cases serve as complementary benchmarks for evaluating the performance 
and scalability of SPHinXsys in handling industrial-scale, 
high-fidelity SPH simulations under distinct computational stress scenarios.
\begin{table}[htbp]
  \centering
  \caption{Comparison of computational efficiency between CPU and GPU simulations}
  \label{table:cpu-gpu-efficiency}
  \small %
  \setlength{\tabcolsep}{5pt} %
  \begin{tabular}{cccccc}
    \toprule
    Case & 
    \makecell{Particles\\(Million)} & 
    \makecell{Time Step\\Size (s)} & 
    \multicolumn{2}{c}{Total Wall Time (h)} & 
    \makecell{Speedup\\Ratio} \\
    \cmidrule(lr){4-5}
    & & & CPU & GPU & \\
    \midrule
    Case 9  & 1.12 & $2.00 \times 10^{-9}$ & 215.76 & 31.2  & 6.92 \\
    Case 10 & 5.70 & $4.00 \times 10^{-8}$ & 1644.0 & 193.2 & 8.50 \\
    \bottomrule
  \end{tabular}
\end{table}

The results, shown in Table \ref{table:cpu-gpu-efficiency}. 
Despite the drastically different bottlenecks the GPU-based solver achieves speedups 
of approximately 7 to 9 times over its CPU counterpart. 
These gains are consistent across different computational demands and confirm 
the robustness of the parallel algorithm under heterogeneous workloads.
Figure \ref{fig:velocity-comparison-cpu-gpu} further compares the internal lubricant 
flow fields at the same simulated moment (after Shaft 3 has completed 5 full revolutions) 
for both cases. The left subfigure (Case 10) shows fine-grained oil structures 
due to higher particle resolution, revealing detailed splash patterns and vortex structures. 
In contrast, the right subfigure (Case 9) captures the extremely high-velocity jetting caused 
by ultra-high rotation, with local oil velocities exceeding 200 m/s, highlighted by the red regions 
in the color map. 
At this operating condition, the characteristic Reynolds number based on tip velocity 
and oil-layer thickness reaches $O(10^3)$, while localized jetting regions may momentarily attain $O(10^4)$. 
This places the flow in a transitional regime, where large-scale structures are well resolved but 
small-scale turbulent motions may not be fully captured by the present iLES formulation.
Under such highly dynamic conditions, 
this wide velocity range introduces considerable numerical stiffness, 
necessitating extremely fine time discretization.

\begin{figure}[htbp]
  \centering
  \begin{subfigure}[b]{0.48\textwidth}
    \includegraphics[width=\textwidth]{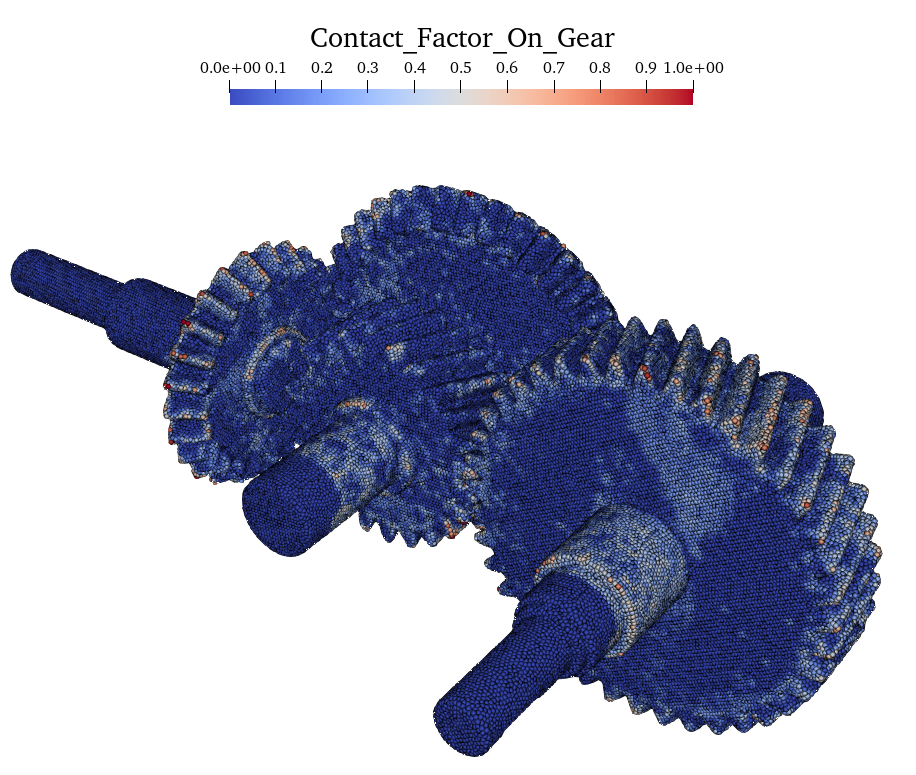}
    \caption{}
    \label{fig:contact factor case 9 N10}
  \end{subfigure}
  \hfill
  \begin{subfigure}[b]{0.48\textwidth}
    \includegraphics[width=\textwidth]{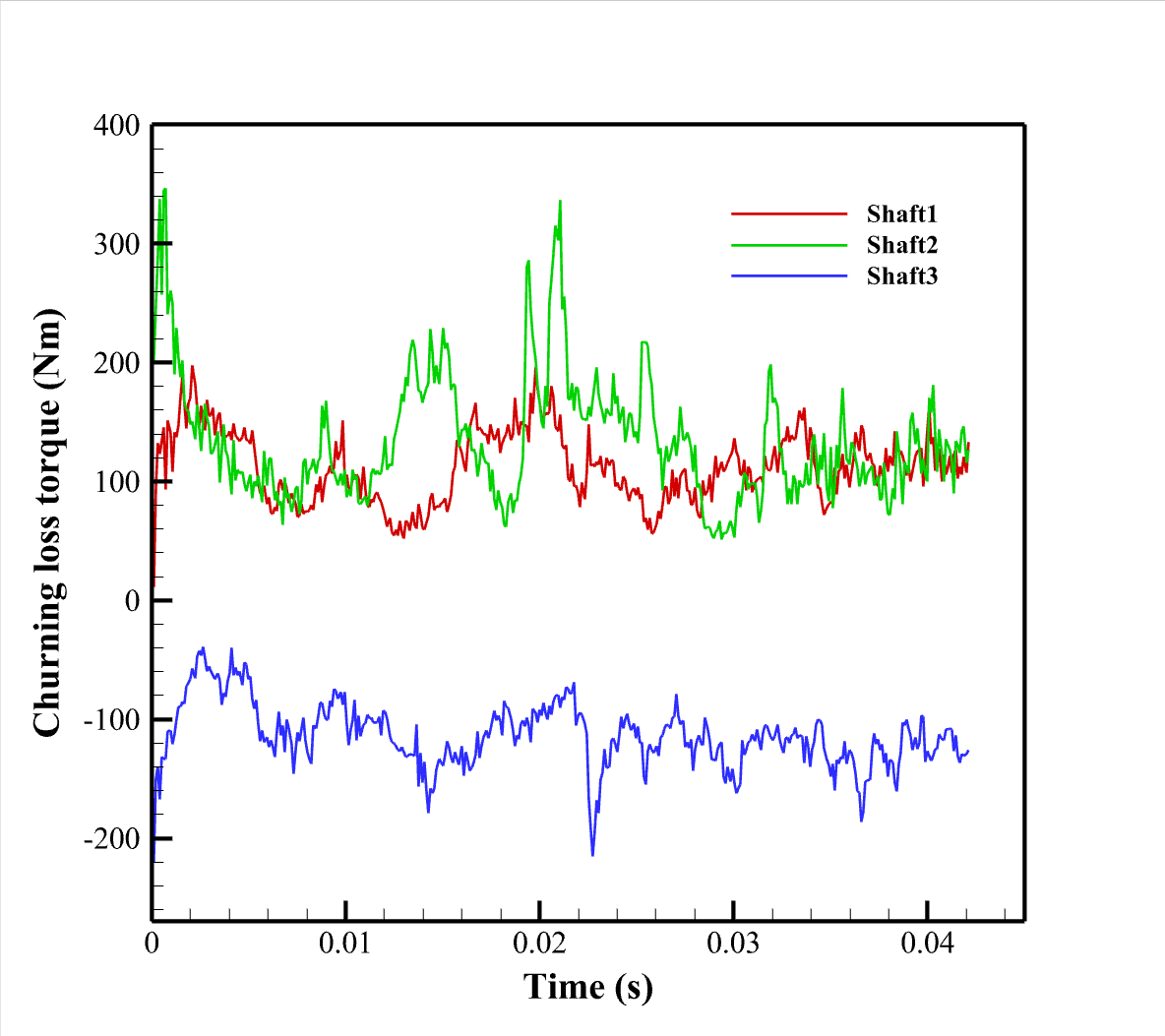}
    \caption{ }
    \label{fig:churning loss case 9}
  \end{subfigure}
  \vspace{0.3cm} 
 \caption{Visualization of simulation results under the extreme high-speed condition 
 (Case 9, $\omega$ = 1500 $\mathrm{rad/s}$). (a) Distribution of Contact factor on the gear surfaces at the moment while the Shaft 3 rotates 10 rounds.
 (b) Time history of churning loss torque for Shaft 1, Shaft 2, and Shaft 3; }
  \label{fig:performance-case-9}
\end{figure}

To complement the performance evaluation under extreme operating conditions, 
Figure \ref{fig:performance-case-9} presents two representative outputs from Case 9 
($\omega$ = 1500 $\mathrm{rad/s}$).
Figure \ref{fig:contact factor case 9 N10} illustrates the contact factor across the gear surfaces, indicating widespread but non-uniform fluid–solid interaction. 
Figure \ref{fig:churning loss case 9} shows the temporal evolution of the churning loss torque 
for each shaft. The results reveal noticeable fluctuations for Shaft 2, while Shaft 1 and Shaft 3
exhibit a relatively stable negative torque, consistent with their high-speed rotation and lower 
immersion level. Although detailed flow or thermal analysis is not shown here, 
these outputs demonstrate that the simulation remains stable and capable of capturing relevant 
physical quantities under highly dynamic conditions.

%
%
%
\section{Concluding remarks}\label{sec:conclusion}

This work presented a GPU-accelerated, fully coupled fluid–solid–thermal SPH solver 
for splash-lubricated gearboxes. The method resolves free-surface lubrication flow, 
fluid–solid interaction, and transient heat transfer within a unified particle 
framework, and introduces a contact-factor metric to quantify instantaneous 
lubricant coverage. Validation against churning-loss measurements of a C-PT FZG gearbox
and an analytical conduction benchmark demonstrated good agreement in both cases.

The multi-parameter study revealed consistent flow and thermal trends. Churning 
losses increase by nearly an order of magnitude as the speed rises from 150 to 
600~$\mathrm{rad/s}$, with oil immersion depth exerting a consistently strong 
influence and viscosity acting as a secondary modifier. Thermal behavior shows 
a similarly clear contrast: 
over the first ten gear rotations, the lubricant temperature rise at 
150~$\mathrm{rad/s}$ is roughly a three to four times larger than at 600~$\mathrm{rad/s}$.
Variations in immersion depth and 
viscosity adjust this heating rate within a modest 10--20\% range, with their 
effects reversing between low- and high-speed regimes. Overall, rotational 
speed dominates both hydrodynamic losses and short-time thermal response.

From a computational standpoint, 
the GPU backend provides a 7–9× speedup over a high-performance desktop CPU, 
even when running on a previous-generation consumer GPU, enabling multi-million-particle, 
long-duration, full-gearbox thermo–fluid simulations without specialized hardware.

Future work will incorporate frictional heat generation at tooth contacts and 
turbulence-enhanced heat transfer to extend the model’s applicability to 
higher-speed conditions.

Overall, the proposed framework provides a practical and high-fidelity tool for 
studying lubricant transport, splash dynamics, churning losses, and thermal evolution 
in industrial gearboxes, and forms a foundation for future work on lubrication and thermal optimization.
%
%
 
%
%
\section*{CRediT authorship contribution statement}
{\bfseries  Yongchuan Yu:} Investigation, Methodology, Visualization, Validation, Formal analysis, Writing - original draft, Writing - review \& editing;
{\bfseries  Dong Wu:} Investigation, Methodology, Validation, Formal analysis, Writing - review \& editing;
{\bfseries  Xiangyu Hu:} Investigation, Methodology, Supervision, Writing - review \& editing;
{\bfseries  Oskar J. Haidn:} Investigation, Supervision, Writing - review \& editing.
%
%
\section*{Declaration of competing interest }
The authors declare that they have no known competing financial interests 
or personal relationships that could have appeared to influence the work reported in this paper.
%
%
\clearpage
\bibliographystyle{elsarticle-num}
\bibliography{gearbox}
%
%
\end{document}